\documentclass[useAMS,usenatbib,usegraphics]{mn2e}
 \usepackage{times}
 \usepackage{amssymb}
 \usepackage{graphicx}
 \usepackage{natbib}

 \title[NGC 1252, a high altitude open cluster remnant] 
       {NGC 1252: a high altitude, metal poor open cluster remnant\thanks{Based on 
        observations collected at the European Organization for 
        Astronomical Research in the Southern Hemisphere, Chile
        (program IDs 281.D-5054 and 086.D-0963).}
       }

 \author[R. de la Fuente Marcos et al.]
        {R.~de~la~Fuente~Marcos,$^{1}$\thanks{E-mail: rfm2013@live.com} \
         C.~de~la~Fuente~Marcos,$^1$ \
         C. Moni Bidin,$^2$ \
         G.~Carraro$^{3, 4}$ \newauthor
         and
         E. Costa$^5$ \\
         $^1$Universidad Complutense de Madrid,
             Ciudad Universitaria, E-28040 Madrid, Spain \\
         $^2$Instituto de Astronom\'{\i}a, Universidad Cat\'olica del Norte,
             Av. Angamos 0610, Antofagasta, Chile \\
         $^3$European Southern Observatory, Alonso de Cordova 3107, 
             Vitacura, Santiago, Chile \\
         $^4$Dipartimento di Astronomia, Universit\`a di Padova, 
             Vicolo dell'Osservatorio 2, I-35122 Padova, Italy \\
         $^5$Departamento de Astronom\'{\i}a, Universidad de Chile,
             Casilla 36-D, Santiago, Chile
        }
 \date{Accepted 2013 June 5.
       Received 2013 May 21;
       in original form 2013 January 9}
 \pubyear{2013}
 
\begin{document}
  \maketitle

  \begin{abstract}
     If stars form in clusters but most stars belong to the field, 
     understanding the details of the transition from the former to 
     the latter is imperative to explain the observational properties 
     of the field. Aging open clusters are one of the sources of field 
     stars. The disruption rate of open clusters slows down with age 
     but, as an object gets older, the distinction between the 
     remaining cluster or open cluster remnant (OCR) and the 
     surrounding field becomes less and less obvious. As a result, 
     finding good OCR candidates or confirming the OCR nature of some 
     of the best candidates still remains elusive. One of these objects 
     is NGC~1252, a scattered group of about 20 stars in Horologium.  
     Here we use new wide-field photometry in the $UBVI$ pass-bands, 
     proper motions from the Yale/San Juan SPM~4.0 catalogue and high 
     resolution spectroscopy concurrently with results from $N$-body 
     simulations to decipher NGC~1252's enigmatic character. 
     Spectroscopy shows that most of the brightest stars in the 
     studied area are chemically, kinematically and spatially 
     unrelated to each other. However, after analysing proper motions, 
     we find one relevant kinematic group. This sparse object is 
     relatively close ($\sim$1~kpc), metal poor and is probably not 
     only one of the oldest clusters (3~Gyr) within 1.5~kpc from the 
     Sun but also one of the clusters located farthest from the disc, 
     at an altitude of nearly -900~pc. That makes NGC~1252 the first 
     open cluster that can be truly considered a high Galactic 
     altitude OCR: an unusual object that may hint at a star formation 
     event induced on a high Galactic altitude gas cloud. We also 
     conclude that the variable TW~Horologii and the blue straggler 
     candidate HD~20286 are unlikely to be part of NGC~1252. 
     NGC~1252~17 is identified as an unrelated, Population II 
     cannonball star moving at about 400~km~s$^{-1}$.
  \end{abstract}


  \section{Introduction}
    It is now widely accepted that stars are formed in some type of star cluster \citep[see e.g.][]{Hopkins13} but the vast majority of 
    stars in any given galaxy are field stars. Understanding the details of the processes responsible for the transition of stars from star 
    clusters (globular and open clusters, stellar associations or any other similar structure) to the field is essential to explain the 
    observed astrometric, kinematic and chemical properties of the field populations. In the Milky Way disc, open star clusters born within 
    star-forming complexes are one of the sources of field stars \citep[see e.g.][]{Elmegreen77,Efremov78,Efremov79,Elmegreen96,Efremov98}. 
    In the absence of a catastrophic encounter with a giant molecular cloud (GMC), theory shows that an open cluster soon starts losing 
    members to the field; first mainly by ejection, then by the most gradual process of evaporation. Early numerical simulations \citep{vHo63} 
    found that open cluster dissolution proceeds slowly at first, then increasingly faster only to slow down when the object is barely 
    distinguishable from the surrounding star field. This intrinsic lack of contrast against the stellar background makes observational 
    studies of the final stages of the evolution of open clusters particularly challenging, especially in the case of the most dynamically 
    evolved and therefore more interesting objects. This may explain why the final stage of the evolution of star clusters has traditionally 
    received more attention from theoretical grounds than from the observational community. Years before the invention of computers, this 
    topic had already attracted the interest of \citet{Ross28}, \citet{Ambar38} and \citet{Spitzer40}. In particular, the seminal papers of 
    \citet{Ambar38} and \citet{Spitzer40} showed that star clusters cannot evaporate completely and that evaporation proceeds only until a 
    hierarchical multiple system is formed \citep{Spitzer40}.  
    
    The almost final residue of the evolution of an open cluster is often called an open cluster remnant or OCR \citep[e.g.][]{dlFM98}. 
    From an observational perspective, OCRs are expected to consist of just a few tens of stars. This can be naively interpreted as a strong 
    argument in favour of a scenario in which such an object should disintegrate (evaporate) within a relatively short time-scale of the 
    order of $t_{\rm relax} \approx (0.1\ N/\ln N) \ t_{\rm cross}$ \citep[actually, $t_{\rm evap} \simeq 140 \ t_{\rm relax}$, see 
    e.g.][]{BT08}, where $t_{\rm relax}$ is the cluster relaxation time, $N$ is the cluster population and $t_{\rm cross}$ is the cluster 
    crossing time or characteristic time-scale required for a star in the open cluster to travel a distance equal to the size of the cluster. 
    Following this (erroneous) line of thought, OCRs should be virtually impossible to observe due to their intrinsically short lives. This 
    reasoning is quite correct in the case of open clusters born that way, i.e. with just a few tens of stars. However, currently 
    observable OCRs likely had many more stars in the past, perhaps as many as $N \sim 10^4$ members when they were born. At the solar 
    circle, an open cluster that rich is expected to last several Gyr. In this case, numerical simulations predict a remnant consisting of 
    mainly binaries and long-lived hierarchical triples \citep[e.g.][]{dlFM98}. For the oldest systems, this population of binaries and 
    triples is the result of several Gyr of close and distant gravitational interactions and also stellar evolution. Therefore, these are 
    not primordial systems but the survivors of a long and highly selective contest for dynamical stability where stellar mass loss has now 
    a very limited role. Hence, the resulting multiple system sample is highly biased towards very stable dynamical configurations. In 
    general, a poorly populated young open cluster and an (perhaps much older) OCR may look similar in terms of membership but, dynamically 
    speaking, they are worlds apart \citep[see the review in e.g.][]{dlFM02}.   
    
    Traditionally ignored by the observational community for the reasons pointed out above, an early approach to the topic was considered by 
    \citet{LR74} and implemented by Lod\'en \citep[see][and references therein]{Loden93}. From this humble beginning, the observational 
    interest in OCRs has been steadily increasing in recent years \citep[e.g.][]{BSDDOP01,Carraro02,Carraro06,PBAC03,VCFS04,BB05,CFVMFBS07,MFFC10,PKBM11}. 
    Besides their role in feeding the field populations and in the validation of computer models, the study of OCRs can also help to explain 
    the existence of very wide binary systems \citep{MC11,RM12}, higher order hierarchical systems \citep{vdBPZMc07} and their high fraction 
    in the field with respect to what is observed in young clusters \citep{KGPDMK10}. Therefore, OCRs are not mere curiosities and, in fact, 
    there are multiple good reasons for studying them. Unfortunately, finding good OCR candidates or, even better, confirming the OCR nature 
    of some of them still remains elusive. Chance alignments of bright stars or asterisms often mimic the structural properties of OCRs 
    \citep[see e.g.][]{MFFC10} and both kinematic and spectroscopic information are generally required to reveal the true nature of these 
    interesting but challenging objects.  
    
    In this paper, we re-examine the possible open cluster remnant nature of NGC~1252, a sparse southern high Galactic latitude group of 
    stars in Horologium. This controversial object has been classified both as asterism and OCR, depending on the authors (see below). The 
    main objective of this research is to shed new light on this matter by providing a detailed study of the stellar content and kinematics 
    of the object. To this end, we present new CCD $UBVI$ photometry and two-epoch spectroscopy, which we combine with proper motions from 
    the Yale/San Juan Southern Proper Motion SPM 4.0 (SPM4) catalogue \citep{GvA11} in an attempt to unravel the true nature and current 
    dynamical status of this object, if real. This paper is organized as follows. In Section 2 we provide an extensive review of the data 
    previously available on this object and its associated area of the sky. Observations, data reduction and overall results are presented 
    in Section 3. In Section 4 we focus on proper motions. The relevance of the results is discussed in Section 5. In Section 6 we draw our 
    conclusions. Some objects are further analysed in Appendix A. 
 
  \section{NGC 1252: a controversial object}
\begin{figure}
     \centering
      \includegraphics[width=\columnwidth]{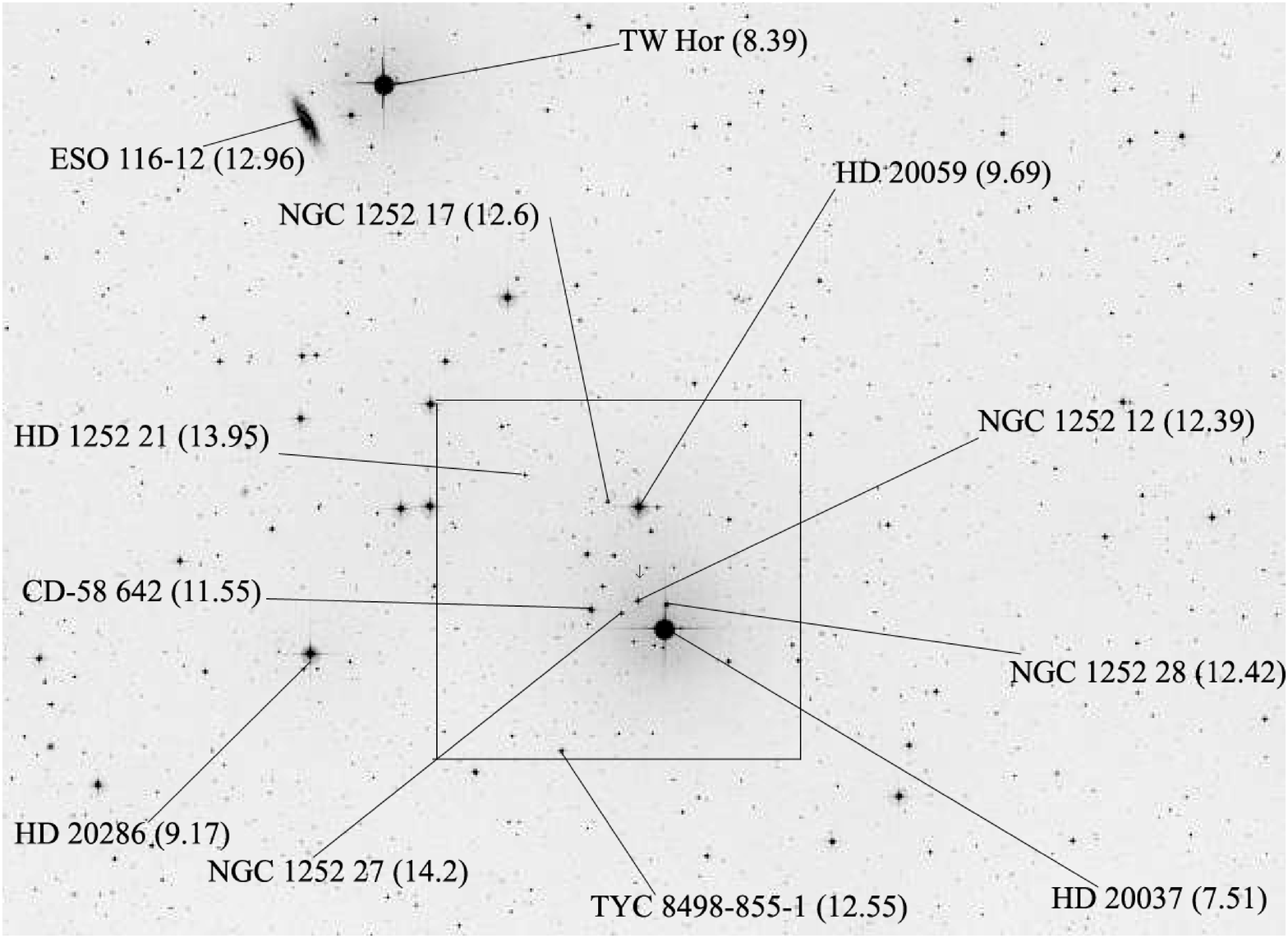}
      \caption{SERC J-DSS1 image of the neighbourhood of NGC~1252 (Epoch 1975-09-10T18:07:00, $\lambda$ = 468~nm, resolution 
               6.8 arcsec pixel$^{-1}$), north is up, east to the left. A number of objects mentioned in the text are indicated, 
               with $B$ magnitudes in parentheses. Within the square, we have the observed $20\times20$ arcmin$^2$ stellar 
               field (see the text for details). The original description in \citet{Dreyer88}, an 8th mag star surrounded by 
               a group of 18 or 20 stars, matches well what is actually observed if we assume that HD~20037 is part of the 
               object.
              }
      \label{ngc1252field}
    \end{figure}

    This sparse group of stars in Horologium was first recorded by Herschel in 1834 as h~2515 \citep{Herschel47}. The same group of stars 
    was included as object GC~663 by \citet{Herschel64} in his `Catalogue of Nebulae and Clusters of Stars'. The `New General Catalogue of 
    Nebulae and Clusters of Stars' \citep{Dreyer88} coined the name NGC~1252 and described the object as an 8th mag star surrounded by a 
    group of 18 or 20 stars (see Fig. \ref{ngc1252field}). The object did not receive further attention until the publication of `The 
    Revised New General Catalogue of Nonstellar Astronomical Objects' \citep{ST73}. This catalogue is a modern, revised and expanded 
    version of the original NGC. Besides incorporating the many corrections to the NGC found over the years, each object was verified on 
    Palomar Observatory Sky Survey (POSS) prints and on plates for southern objects specifically taken for the purpose. In this catalogue, 
    NGC~1252 is described as an unverified southern object, i.e. the object is regarded as doubtful. The Catalogue of Star Clusters and 
    Associations \citep{RBW81} does not include this object. 
%
%
%
%

    An extensive search across published literature on this object reveals over a dozen entries, several of them within the context of the 
    OCR paradigm. The first detailed study of NGC~1252 was carried out by \citet{Bouchet83}. They found that NGC~1252 is an open cluster of 
    age $\sim$500 Myr located at a distance of nearly 470 pc with a diameter of nearly $0\fdg5$ or 8 pc. They also pointed out that the 
    carbon star TW~Horologii (= HD~20234) is probably a member of the cluster (see Fig. \ref{ngc1252field} for their relative positions). In 
    contrast, \citet{Eggen84} considered the existence of such a cluster unlikely and listed the variable star TW~Horologii, a putative 
    member of NGC~1252 for the previous authors, as a probable member of the Hyades supercluster. \citet{Ahumada95} identified a blue 
    straggler candidate (HD~20286) in NGC~1252 (see Fig. \ref{ngc1252field}), that they considered an open cluster based on \citet{Bouchet83} 
    data. More recently, \citet{Baumgardt98} and Baumgardt, Dettbarn \& Wielen (2000) concluded that NGC~1252 is not a real cluster but an 
    asterism. Feeding the controversy, \citet{Pavani01} and \citet{BSDDOP01} argued that NGC~1252 is not a field fluctuation but an OCR of 
    age 3 Gyr located 640 pc from the Sun and 460 pc below the Galactic disc. \citet{LB03} also considered NGC~1252 an open cluster located 
    707 pc from the Sun and provided proper motions (see below) based on six candidate members. Following \citet{Bouchet83} and 
    \citet{Pavani01}, \citet{XD05} also described NGC~1252 as an extremely underpopulated open cluster with a very luminous blue straggler, 
    HD~20286. Using United States Naval Observatory CCD Astrograph Catalogue 2.0 \citep[UCAC2;][]{ZUZ04} positions and proper motions, 
    \citet{Dias06} classified NGC~1252 as open cluster with proper motions ($\mu_{\alpha} \cos\delta, \mu_{\delta}) = (14.28, 9.07)$ mas 
    yr$^{-1}$. They estimated the number of members in the cluster at 25. Based on \citet{Pavani01} and \citet{Dias02}, \citet{vdBergh06} 
    gives a value of 2.61 pc for the diameter of the cluster. \citet{PB07} reanalysed Two Micron All Sky Survey \citep[2MASS;][]{Skru97} 
    photometry and UCAC2 proper motions to confirm NGC~1252 as a 2.8 Gyr old loose OCR located at a distance of 790 pc from the Sun and 610 
    pc below the disc. \citet{Ahumada07} ratified their 1995 results using data from \citet{Dias02}. \citet{PKBM11} compiled previous 
    results to conclude that NGC~1252 is a robust OCR candidate. \citet{ZPBML12} also consider NGC~1252 among open clusters, with proper 
    motions ($\mu_{\alpha} \cos\delta, \mu_{\delta}) = (11.1, 7.5)$ mas yr$^{-1}$. 
    
    NGC~1252 (also known as ESO~116--11) is currently classified as a cluster of stars in SIMBAD.\footnote{http://simbad.u-strasbg.fr} Its 
    FK5 coordinates are given as $\alpha=03^{\rm h}~10^{\rm m}~49^{\rm s}$, $\delta=-57\degr 46\arcmin 00\arcsec$ \citep{XD05} with Galactic 
    coordinates $l=274\fdg084$, $b=-50\fdg831$. Its proper motion is cited as ($\mu_{\alpha} \cos\delta, \mu_{\delta}) = (7.98\pm0.60, 
    5.95\pm0.45)$ mas yr$^{-1}$ \citep{LB03}. NGC~1252 is listed as a {\it bona fide} open cluster in both the Open Cluster 
    Database\footnote{http://www.univie.ac.at/webda/} \citep[WEBDA;][]{MP03} and the New Catalogue of Optically Visible Open Clusters 
    and Candidates\footnote{http://www.astro.iag.usp.br/$\sim$wilton/} \citep[NCOVOCC;][]{Dias02}. These data bases are widely used in 
    professional open cluster studies. The current version of WEBDA \citep[as of March 2013;][]{PM13} includes coordinates identical to 
    those in SIMBAD, distance, 640 pc, reddening, 0.02 mag, distance modulus, 9.09 mag, age, 3.0 Gyr, and diameter, 14 arcmin. Therefore, 
    the object is located 496 pc below the Galactic plane. This is rather unusual for an open cluster and it will be discussed at length 
    later. The January 2013 version \citep[v3.3;][]{Dias13} of NCOVOCC includes NGC~1252 as an OCR \citep[based on][]{PB07} with coordinates 
    identical to those in SIMBAD and diameter, 8 arcmin, located 790 pc from the Sun (therefore, 612 pc below the Galactic plane), its 
    colour excess is listed as 0.00 mag with an age of 2.8 Gyr and proper motions ($\mu_{\alpha} \cos\delta, \mu_{\delta}) = (7.98\pm0.45, 
    5.95\pm0.60)$ mas yr$^{-1}$ \citep[][note the switching of errors with respect to the SIMBAD data above]{LB03}. The NGC/IC 
    Project\footnote{http://www.ngcicproject.org} \citep{Erdmann10}, a source popular among amateur astronomers, indicates that NGC~1252 is 
    a sparsely populated cluster of size 2 arcmin \citep[see also][]{Corwin04}.

    The roots of the debate on the nature of NGC~1252 can be traced back to an issue frequently overlooked: despite being an object located 
    at high Galactic latitude and, in theory, mostly free from stellar contamination, the area of the sky around NGC~1252 is also home of 
    two relatively close and well-studied stellar structures, the Tucana-Horologium association (e.g. Zuckerman, Song \& Webb 2001) and the 
    Hyades stream or supercluster \citep[e.g.][]{Egg59} and, perhaps, additional relatively close moving groups (fully evaporated open 
    clusters not OCRs). The Tucana-Horologium association contains a coeval ($\sim$30 Myr old) stream of stars with common space motion all 
    within $\sim$70 pc from the Earth \citep{ZRSB11}. Members of the Tucana-Horologium association in the area of NGC~1252 have heliocentric 
    distances $\sim$50 pc, $V$ in the range 4-16 mag and proper motions $\sim$70 mas yr$^{-1}$. Only a few of the brightest stars in our 
    samples are expected to be part of the Tucana-Horologium association. Far more problematic could be the role of the Hyades stream which 
    is expected to be a main source of stellar contamination in the field of NGC~1252. The existence of the Hyades stream was first proposed 
    by Olin J. Eggen and it was first studied in detail by \citet{OgLa68} concluding that it contained too many stars to having been 
    supplied by the Hyades cluster alone. The debate on the true nature of the Hyades stream has only recently been settled; instead of 
    being primarily composed of coeval stars originating from the evaporating Hyades open cluster, the structure is believed to be the 
    result of resonant trapping induced by the Galactic spiral perturbation, specifically, scattering at a Lindblad resonance 
    \citep{Fama07,Sell10,HSP11,Pom11}. The Hyades stream contains an heterogeneous group of stars mostly within 500 pc from the Earth, their 
    radial velocities close to 11~km~s$^{-1}$ in the area around NGC~1252 \citep{Eggen84} and they can be included in an imaginary box with 
    Galactic velocity components U $\in$ (-50,~-25)~km~s$^{-1}$ and V $\in$ (-23,~-12)~km~s$^{-1}$ \citep{Pom11}.

  \section{Observations and reduction}
    The observations presented here were completed in three separate runs and focused on the area enclosed by the square in Fig. 
    \ref{ngc1252field}. Photometry was obtained on 2008 December at Cerro Tololo. Spectroscopy was completed in two different runs, both at 
    La Silla, on 2008 September and 2010 December, respectively. Details of the actual observations and the data reduction procedures as 
    well as the results are given in the following sections.  

    \subsection{Photometry}
      The region of interest ($20\times20$ arcmin$^2$ area enclosed by the square in Fig.~\ref{ngc1252field}) was observed with the Y4KCAM 
      camera attached to the Cerro Tololo Inter-American Observatory (CTIO) 1-m telescope, operated by the SMARTS 
      consortium\footnote{http://www.astro.yale.edu/smarts}. This camera is equipped with an STA 4064$\times$4064 
      CCD\footnote{http://www.astronomy.ohio-state.edu/Y4KCam/ detector.html} with 15-$\mu$m pixels, yielding a scale of 
      0.289$\arcsec$/pixel and a field-of-view (FOV) of $20\times20$ arcmin$^2$ at the Cassegrain focus of the CTIO 1-m telescope. The 
      CCD was operated without binning, at a nominal gain of 1.44 e$^-$/ADU, implying a readout noise of 7~e$^-$ per quadrant (this detector 
      is read by means of four different amplifiers). The quantum efficiency and other detector characteristics can be found at the OSU web 
      site.$^{6}$ In Table~\ref{log} we present the log of our $UBVI$ observations. All the observations were carried out in good-seeing, 
      photometric conditions. Our $UBVI$ instrumental photometric system was defined by the use of a standard broad-band Kitt Peak 
      $UBVI_{kc}$ set of filters.\footnote{http://www.astronomy.ohio-state.edu/Y4KCam/filters.html} To determine the transformation from our 
      instrumental system to the standard Johnson-Kron-Cousins system and to correct for extinction, we observed 46 stars in Landolt's area 
      SA~98 \citep{Lan92} multiple times and with different airmasses ranging from $\sim$1.2 to $\sim$1.8. Field SA~98 is very advantageous, 
      as it includes a large number of well-observed standard stars, with a very good colour coverage: $-0.2 \leq (B-V)\leq 2.2$ and $-0.1 
      \leq (V-I) \leq 6.0$. Furthermore, it is completely covered by the FOV of the Y4KCAM. In addition, observations of standard star 
      fields PG~1047, SA~104 and PG~0918 were obtained in order to have an even wider airmass coverage.
%
%
      \begin{table}
        \tabcolsep 0.12truecm
        \caption{$UBVI$ photometric observations of NGC~1252 and standard star fields.}
        \begin{tabular}{lllll}
          \hline
            \noalign{\smallskip}
            Target & Date & Filter & Exposure (s) & Airmass \\
            \noalign{\smallskip}
          \hline
            \noalign{\smallskip}
            SA~98    & 2008 December 25  & \textit{U} & 2$\times$10, 2$\times$150, 2$\times$400   & 1.15--1.55 \\
                     &                   & \textit{B} & 2$\times$10, 2$\times$100, 2$\times$200   & 1.15--1.41 \\
                     &                   & \textit{V} & 2$\times$10, 2$\times$60, 2$\times$120    & 1.15--1.46 \\
                     &                   & \textit{I} & 2$\times$10, 2$\times$60, 2$\times$120    & 1.15--1.35 \\
            NGC~1252 & 2008 December 25  & \textit{B} & 10, 100, 1500                             & 1.27--1.33 \\
                     &                   & \textit{V} & 10, 100, 900                              & 1.34--1.36 \\
                     &                   & \textit{I} & 10, 60, 120, 900                          & 1.21--1.24 \\
            SA~104   & 2008 December 25  & \textit{U} & 30, 3$\times$200                          & 1.61--1.69 \\
                     &                   & \textit{B} & 10, 100                                   & 1.77--1.78 \\
                     &                   & \textit{V} & 10, 100                                   & 1.72--1.74 \\
                     &                   & \textit{I} & 10, 100                                   & 1.82--1.84 \\
            SA~98    & 2008 December 26  & \textit{U} & 2$\times$10, 2$\times$150, 2$\times$400   & 1.15--1.62 \\
                     &                   & \textit{B} & 2$\times$10, 2$\times$100, 2$\times$200   & 1.15--1.81 \\
                     &                   & \textit{V} & 2$\times$10, 2$\times$60, 2$\times$120    & 1.15--1.67 \\
                     &                   & \textit{I} & 2$\times$10, 2$\times$60, 2$\times$120    & 1.15--1.72 \\
            NGC~1252 & 2008 December 26  & \textit{U} & 2$\times$20, 200, 2000                    & 1.15--1.16 \\
                     &                   & \textit{B} & 10, 100                                   & 1.20--1.20 \\
                     &                   & \textit{V} & 10, 60                                    & 1.19--1.20 \\
                     &                   & \textit{I} & 10, 60                                    & 1.21--1.21 \\
            PG~1047  & 2008 December 26  & \textit{U} & 20, 200                                   & 1.22--1.23 \\
                     &                   & \textit{B} & 10, 150                                   & 1.21--1.21 \\
                     &                   & \textit{V} & 10, 60                                    & 1.22--1.22 \\
                     &                   & \textit{I} & 10, 60                                    & 1.20--1.20 \\
            PG~0918  & 2008 December 26  & \textit{U} & 10, 200, 400                              & 2.00--2.11 \\
                     &                   & \textit{B} & 10, 200                                   & 1.77--1.79 \\
                     &                   & \textit{V} & 10, 100                                   & 1.88--1.90 \\
                     &                   & \textit{I} & 10, 100                                   & 1.82--1.83 \\
            \noalign{\smallskip}
          \hline
        \end{tabular}
        \label{log}
      \end{table}
%
%

      \subsubsection{Reductions}
        Basic calibration of the CCD frames was completed using the Yale/SMARTS y4k reduction script based on the \textsc{iraf}\footnote{\textsc{iraf} 
        is distributed by the National Optical Astronomy Observatory, which is operated by the Association of Universities for Research in 
        Astronomy, Inc., under cooperative agreement with the National Science Foundation.} package \textsc{ccdred}. For this purpose, zero 
        exposure frames and twilight sky flats were acquired every night. Photometry was then performed using the \textsc{iraf} \textsc{daophot} 
        and \textsc{photcal} packages. Instrumental magnitudes were extracted following the point spread function (PSF) method \citep{Stet87}. 
        A quadratic, spatially variable, master PSF (PENNY function) was adopted. Aperture corrections were determined making aperture 
        photometry of a suitable number (typically 10 to 20) of bright, isolated, stars in the field. These corrections were found to vary 
        from 0.160 to 0.290 mag, depending on the filter. The PSF photometry was finally aperture corrected, filter by filter.

      \subsubsection{Results}
        After removing problematic stars (blends, possible variable stars, artifacts, etc) and stars having only a few observations (less 
        than 5) in Landolt's (1992) catalogue, our photometric solution for a grand total of 322 measurements per filter, turned out to be:\\ \\
        \noindent
          $ U = u + (3.080\pm0.010) + (0.45\pm0.01) \times X - (0.009\pm0.006) \times (U-B)$ \\
          $ B = b + (2.103\pm0.012) + (0.27\pm0.01) \times X - (0.101\pm0.007) \times (B-V)$ \\
          $ V = v + (1.760\pm0.007) + (0.15\pm0.01) \times X + (0.028\pm0.007) \times (B-V)$ \\
          $ I = i + (2.751\pm0.011) + (0.08\pm0.01) \times X + (0.045\pm0.008) \times (V-I)$ \\ \\
        \noindent
        where $UBVI$ are standard magnitudes, $ubvi$ are the instrumental ones and $X$ is the airmass. The coefficients have been 
        obtained averaging the photometric solutions of individual nights (December 25 and 26) into a single one, since the two photometric 
        solutions turned out to be identical within the observational uncertainties. The final rms of the fitting were 0.030, 0.015, 0.010 
        and 0.010 in $U$, $B$, $V$ and $I$, respectively.
      
\begin{figure}
        \centering
          \includegraphics[width=\columnwidth]{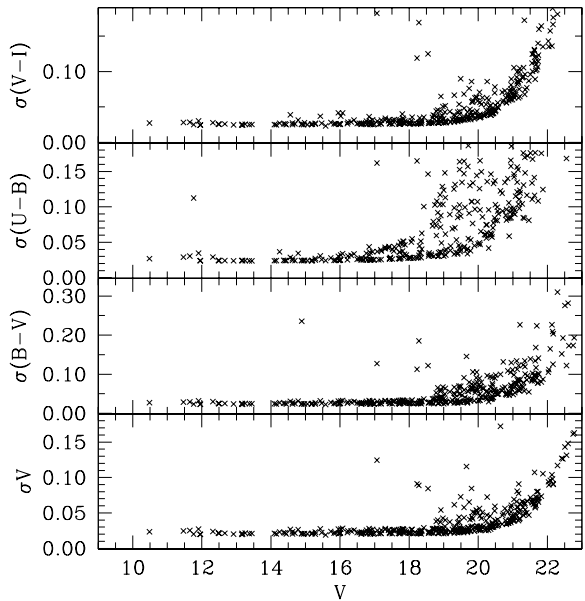}
          \caption{Photometric errors in $V$, $(B-V)$, $(U-B)$ and $(V-I)$ as a function of the $V$ magnitude.}
          \label{errors}
      \end{figure}

        Global photometric errors were estimated using the scheme developed by \citet[][appendix A1]{PC01}, which takes into account the 
        errors resulting from the PSF fitting procedure (i.e. from \textsc{allstar}) and the calibration errors (corresponding to the zero 
        point, colour terms and extinction errors). In Fig.~\ref{errors} we present our global photometric errors in $V$, $(B-V)$, $(U-B)$ 
        and $(V-I)$ plotted as a function of the $V$ magnitude. Quick inspection shows that stars brighter than $V \approx 20$ mag have 
        errors smaller than $\sim0.05$~mag in magnitude and lower than $\sim0.10$~mag in $(B-V)$ and $(V-I)$. Larger errors are found in 
        $(U-B)$. Our final optical photometric catalogue consists of 356 entries having $UBVI$ measurements down to $V\sim 22$ mag.

%
%
%
%

      \subsubsection{Completeness and astrometry}
        Completeness corrections were determined by running artificial star experiments on the data \citep[see][for details]{CBPMS05}. 
        Basically, we created several artificial images by adding artificial stars to the original frames. These stars were added at random 
        positions and had the same colour and luminosity distribution of the true sample. To avoid generating overcrowding, in each 
        experiment we added up to 20 per cent of the original number of stars. Depending on the frame, between 1000 and 5000 stars were 
        added. Then, a systematic comparison between the number of added and detected stars was carried out. In this way, we have estimated 
        that the completeness level of our photometry is better than 50 per cent down to $V = 21.5$, that can be considered as the limiting 
        magnitude of the present study.

        Our optical catalogue was cross-correlated with 2MASS, which resulted in a final catalogue including $UBVI$ and $JHK_{s}$ magnitudes. 
        As a by-product, pixel (i.e. detector) coordinates were converted to RA and Dec. for J2000.0 equinox, thus providing 2MASS-based 
        astrometry.

      \subsubsection{Comparison with previous photometry}
\begin{figure}
     \centering
      \includegraphics[width=\columnwidth]{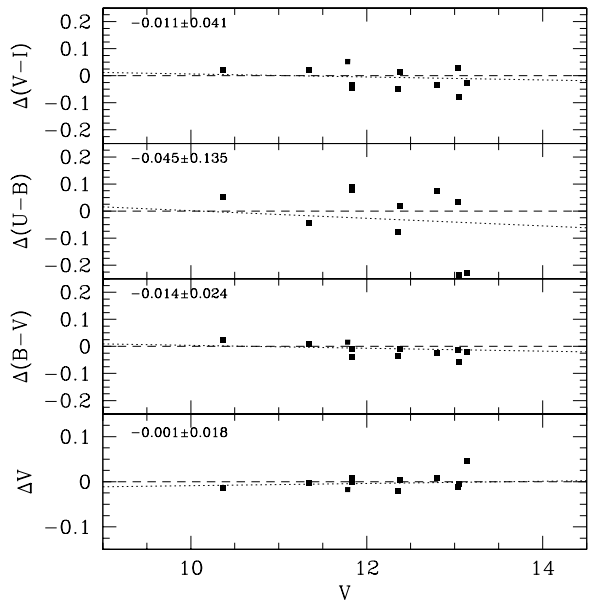}
      \caption{Comparison of our photometry with \citet{Bouchet83} for $V$, $(B-V)$, $U-B$ and $(V-I)$: our photometry minus that in 
               \citet{Bouchet83}. The mean differences are displayed on the top left corners of the various panels.}
      \label{comparison1}
    \end{figure}

        We compared our photometry with that from the $UBVRI$ photoelectric study by \citet{Bouchet83}. We could only compare $UBVI$, since 
        we did not observe in $R$. We found 13 stars in common and the results are shown in Fig.~\ref{comparison1}, in the sense our 
        photometry minus that in \citet{Bouchet83}. We found that the two studies are in the same system, both in $V$ mag, and in all the 
        colours. The mean differences are reported on the top left corners of the various panels in Fig.~\ref{comparison1}. We only notice 
        that a small, unaccounted for, colour term (dashed line) can perhaps be present in $U-B$.

%
%
%
%
  
\begin{figure}
     \centering
      \includegraphics[width=\columnwidth]{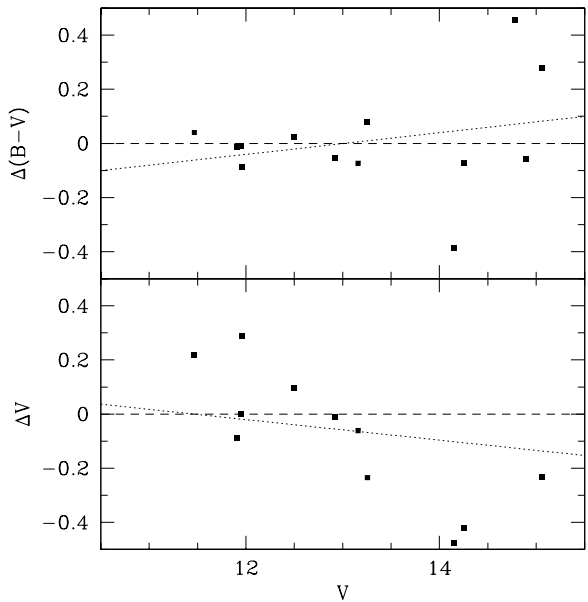}
      \caption{Comparison of our photometry with \citet{Pavani01} for $V$ and $(B-V)$: our photometry minus that in \citet{Pavani01}.
               There are unusually large differences, mainly in $V$. The linear fittings are displayed as dotted lines. Two points fall 
               outside the plotting boundaries at the bottom panel.}
      \label{comparison2}
    \end{figure}

        We then compare our data with the more recent CCD study in $BV$ by \citet{Pavani01}, with which we have 11 stars in common. The 
        comparison is shown in Fig.~\ref{comparison2}, in the sense our photometry minus that in \citet{Pavani01}. There are unusually large 
        differences, mainly in $V$. We do not have a clear explanation for that, since \citet{Pavani01} claim that they used stars in common 
        with \citet{Bouchet83} to tie their photometry. However, a quick glance at their table~1 immediately shows that there are 
        significant differences between them and \citet{Bouchet83} too. The enormous colour term in $V$ simply demonstrates that their 
        photometry must be, somehow, wrong. 

%
%
%
%

      \subsubsection{Colour--magnitude diagrams} 
\begin{figure}
     \centering
      \includegraphics[width=\columnwidth]{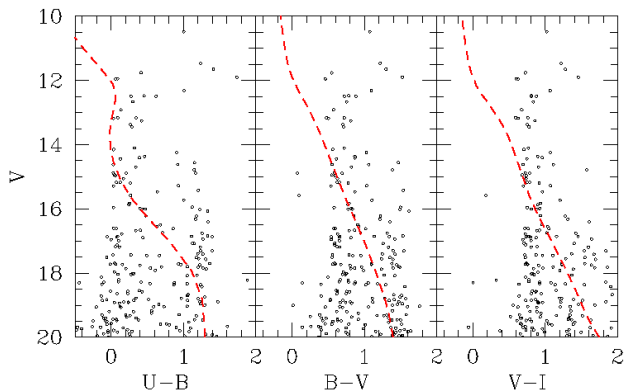}
      \caption{CMDs in various colour combinations for all the stars having $UBVI$ photometry in the field of NGC~1252.
               A ZAMS of solar metallicity taken from the Padova data base \citep{GBBC00} is also included. E($B-V$)=0.02 
               and a distance of 1 kpc have been assumed.
              }
      \label{ngc1252cmd}
    \end{figure}

        From our photometric data we construct colour--magnitude diagrams (CMDs) for several colour combinations and they are shown in Fig. 
        \ref{ngc1252cmd}. The CMDs for the detected stars do not exhibit an obvious, distinctive feature that can lead us to conclude that 
        there is a normal open cluster in the area of the sky commonly assigned to NGC~1252. As a guidance, a zero-age main-sequence (ZAMS) 
        of solar metallicity taken from the Padova data base \citep{GBBC00} is also included. A well populated sequence in the $V$ versus 
        $B - V$ and $V$ versus $V - I$ diagrams can easily be discarded. However, the existence of a real stellar grouping at the location 
        of NGC~1252 cannot be confirmed or ruled out on the basis of photometric evidence alone. 
%
%
%
%

    \subsection{Spectroscopy}
      High-resolution spectra of 13 cluster member candidates were collected at La Silla Observatory during two observing runs. In the
      following, 12 observed stars will be identified by means of their ID in the \citet{Bouchet83} photometry, while the star 
      GSC~0849800928, not included in that catalogue, will be simply referred to as `star~A'. In other words, if a star appears as 
      NGC~1252~1 in SIMBAD, it will be named here as BT1. The spectra of nine stars were secured with the High Accuracy Radial velocity 
      Planet Searcher (HARPS) fibre-fed spectrograph at the 3.6m telescope on 2008 September 18. The details on the actual observations and 
      data reduction methodology were provided in \citet{MFFC10}. A second epoch for six promising candidates, plus the spectrum of four 
      stars not previously observed, were collected with The Fiber-fed Extended Range Optical Spectrograph (FEROS) at the ESO/MPG 2.2m 
      telescope, on 2010 December 17--22. Exposure times varied between 600s and 1800s, depending on target magnitude, for a resulting 
      signal-to-noise ratio (S/N) of 30--60, except for one spectrum of much poorer quality (S/N=10). The spectra were de-biased, 
      flat-fielded, extracted and wavelength calibrated by means of standard \textsc{iraf}$^{8}$ routines. This procedure did not improve 
      the spectral quality with respect to the output of the online pipeline at the telescope, which was therefore used in the analysis. The 
      second fibre of the spectrograph was allocated to the sky during the observations, and the extracted sky background was subtracted from 
      each target.

%
%
      \begin{table}
       \tabcolsep 0.09truecm
       \begin{center}
        \caption{Stellar data derived from spectroscopic analysis}
        \label{t_respec}
        \begin{tabular}{cccccccc}
         \hline
         ID & Run & T$_\mathrm{eff}$ & $\log(g)$ & [$\frac{\mathrm{Fe}}{\mathrm{H}}$] & $\xi$         & RV            & d  \\
            &     & (K)              & (dex)     & (dex)                              & (km s$^{-1}$) & (km s$^{-1}$) & (pc) \\
         \hline
          BT11 & FEROS & 4750 & 3.15 & +0.01   & 1.72 & $-$7.1$\pm$0.6  & 380$\pm$100  \\
          BT12 & FEROS & 6090 & 3.55 & $-$0.50 & 1.83 & 21.8$\pm$0.6    & 1000$\pm$240 \\
               & HARPS & 6030 & 3.65 & $-$0.55 & 1.71 & 22.0$\pm$0.8    & 830$\pm$160  \\
          BT14 & FEROS & 5890 & 4.40 & 0.00    & 1.01 & 31.0$\pm$0.5    & 450$\pm$100  \\
               & HARPS & 5860 & 4.35 & $-$0.05 & 1.04 & 31.1$\pm$0.5    & 460$\pm$130  \\
          BT15 & HARPS & 4360 & 1.65 & $-$0.75 & 1.51 & 8.6$\pm$0.5     & 2140$\pm$370 \\
          BT16 & FEROS & 4850 & 4.45 & +0.05   & 0.61 & 41.1$\pm$0.5    & 200$\pm$20   \\
               & HARPS & 4760 & 4.25 & $-$0.05 & 0.63 & 40.9$\pm$0.3    & 190$\pm$25   \\
          BT17 & HARPS & 4040 & 0.55 & $-$1.15 & 1.79 & 67.7$\pm$0.5    & 6000$\pm$500 \\
          BT18 & FEROS & 5400 & 3.70 & $-$0.30 & 1.48 & 18.4$\pm$0.5    & 770$\pm$200  \\
               & HARPS & 5340 & 3.50 & $-$0.45 & 1.62 & 18.3$\pm$0.5    & 960$\pm$240  \\
          BT19 & FEROS & 5210 & 4.30 & $-$0.12 & 0.88 & 14.3$\pm$0.4    & 300$\pm$50   \\
               & HARPS & 5120 & 4.35 & $-$0.10 & 0.78 & 14.5$\pm$0.5    & 250$\pm$50   \\
          BT21 & FEROS & 5470 & 4.40 & +0.11   & 0.85 & 20.3$\pm$0.6    & 360$\pm$40   \\
          BT22 & FEROS & 5320 & 4.40 & +0.15   & 0.82 & $-$1.4$\pm$0.6  & 240$\pm$30   \\
          BT27 & FEROS & 6040 & 3.75 & $-$0.35 & 1.62 & 5.2$\pm$0.4     & 1200$\pm$300 \\
               & HARPS & 6180 & 3.65 & $-$0.20 & 1.76 & 5.3$\pm$0.5     & 1600$\pm$400 \\
          BT28 & FEROS & 6050 & 3.40 & $-$0.35 & 1.95 & $-$4.0$\pm$0.4  & 1300$\pm$300 \\
          A    & HARPS & 5900 & 4.25 & $-$0.25 & 1.10 & $-$14.5$\pm$0.4 & 680$\pm$60   \\
         \hline
          $\sigma$ &   & 85   & 0.15 & 0.07    & 0.25 &                 &              \\
         \hline
        \end{tabular}
       \end{center}
      \end{table}
%
%
%
%
      \begin{table*}
       \tabcolsep 0.3truecm
       \begin{center}
        \caption{SPM4 proper motions and computed kinematic data. $N$ is the number of stars with proper motions within 3$\sigma$ of those 
                 of a given star in the table (within a $20\times20$ arcmin$^2$ field around NGC~1252). U, V, W are the components of the 
                 heliocentric Galactic velocity of the star (see the text for details).}
        \label{kin}
        \begin{tabular}{ccccccccccc}
         \hline
          ID   & $\alpha$   & $\delta$                    & $V$  & SPM4 ID    & $\mu_{\alpha} \cos\delta$ & $\mu_{\delta}$ & $N$& U          & V          & W \\
               & $(^{\rm h}:^{\rm m}:^{\rm s})$  & ($\degr$:$\arcmin$:$\arcsec$) & (mag)  &            & (mas yr$^{-1}$)  & (mas yr$^{-1}$)  &    & (km s$^{-1}$) & (km s$^{-1}$) & (km s$^{-1}$) \\
         \hline
          BT11 & 03:11:09.0 & -57:47:38                   & 10.41 & 1230000671 & 17.4$\pm$1.5   & 15.9$\pm$1.3   & 2  & -41$\pm$9   &  -5$\pm$7  &  11$\pm$6 \\
          BT12 & 03:10:50.1 & -57:47:09                   & 12.03 & 1230015455 &  0.7$\pm$1.4   &  2.7$\pm$1.3   & 4  & -10$\pm$6   & -11$\pm$5  & -20$\pm$4 \\
          BT14 & 03:11:04.4 & -57:46:23                   & 12.43 & 1230015464 &-11.1$\pm$1.4   &-21.5$\pm$1.3   & 2  &  53$\pm$3   & -22$\pm$2  & -19$\pm$2 \\
          BT15 & 03:11:10.7 & -57:44:38                   & 11.44 & 1230000682 &  3.0$\pm$1.4   &  1.7$\pm$1.2   & 4  & -31$\pm$13  & -19$\pm$11 &   2$\pm$9 \\
          BT16 & 03:10:59.9 & -57:44:42                   & 13.06 & 1230083288 & 23.8$\pm$1.8   & -0.8$\pm$1.7   & 7  & -10$\pm$2   & -41$\pm$1  & -20$\pm$1 \\
          BT17 & 03:11:02.6 & -57:41:49                   & 12.29 & 1230015463 &  7.3$\pm$1.4   &-17.5$\pm$1.3   & 5  & 305$\pm$55  &-380$\pm$40 & 239$\pm$33\\
          BT18 & 03:10:44.9 & -57:43:24                   & 12.64 & 1230015452 &-19.6$\pm$1.6   &-46.4$\pm$1.5   & 1  & 204$\pm$7   & -35$\pm$5  &  16$\pm$4 \\
          BT19 & 03:10:42.3 & -57:42:07                   & 13.25 & 1230083276 &-16.8$\pm$1.8   &-54.8$\pm$1.7   & 2  &  72$\pm$6   & -23$\pm$4  &   4$\pm$3 \\
          BT21 & 03:11:36.1 & -57:40:23                   & 12.64 & 1230015481 & 13.4$\pm$1.7   &  7.1$\pm$1.6   & 14 & -22$\pm$4   & -23$\pm$3  &  -9$\pm$2 \\
          BT22 & 03:11:45.4 & -57:37:48                   & 12.51 & 1230015490 & -2.1$\pm$1.5   & -6.1$\pm$1.3   & 5  &   7$\pm$2   &   0$\pm$1  &   3$\pm$1 \\
          BT27 & 03:10:56.6 & -57:47:48                   & 14.33 & 1230015460 &  6.8$\pm$1.5   & 13.2$\pm$1.4   & 5  & -98$\pm$16  &   2$\pm$11 & -14$\pm$9 \\
          BT28 & 03:10:38.5 & -57:47:20                   & 12.15 & 1230015448 &  9.6$\pm$1.4   &  7.4$\pm$1.3   & 7  & -71$\pm$15  & -19$\pm$12 &  16$\pm$10\\
          A    & 03:10:51.6 & -57:49:21                   & 14.33 & 1230015456 & -7.6$\pm$1.9   &-22.5$\pm$1.9   & 2  &  73$\pm$9   &  -4$\pm$6  &  26$\pm$5 \\
         \hline
        \end{tabular}
       \end{center}
      \end{table*}
%
%

      The radial velocities (RVs) were measured by means of a cross-correlation technique \citep{Tonry79}, as implemented in the 
      \textsc{fxcor} \textsc{iraf} task. A synthetic spectrum of solar metallicity extracted from the library of \citet{Coelho05} was used 
      as template. Its temperature was fixed as the value of the grid nearest to the temperature of the target, estimated from its ($B-V$) 
      colour by means of the relations in \citet{Alonso99}. The typical surface gravity of a red giant, subgiant or main-sequence star 
      ($\log(g)$=2.0, 3.5, 4.5, respectively) was adopted after determining the most likely luminosity class of the target by means of a 
      visual inspection of its spectra. Previous investigations have shown that even relatively large mismatches between the parameters of 
      the object and the synthetic template have only negligible effects on the measurements, although they increase the estimated uncertainties 
      \citep{Morse91,MFFC10,Moni11}. The spectral range 4800--6800~\AA\ was used in the cross-correlation, excluding the 5250--5350~\AA\ gap 
      in the HARPS spectra. The results were corrected for heliocentric velocity, but no other corrections were applied. In fact, zero-point 
      errors were found negligible in HARPS spectra \citep[see discussion in][]{MFFC10} and the excellent agreement between the results 
      obtained in the two epochs indicated that FEROS data neither required correction. The results are given in Table~\ref{t_respec}. 
      The errors were estimated quadratically summing the contribution of the most relevant sources: the cross-correlation error 
      (0.3--0.8~km~s$^{-1}$), the uncertainty in zero-point definition (0.1~km~s$^{-1}$) and choice of synthetic template (0.2~km~s$^{-1}$). 
      The six stars observed in both epochs show no RV variation and the two measurements agree within 0.2~km~s$^{-1}$; this suggests that 
      they are not binaries.

      The stellar parameters (effective temperature, surface gravity and microturbulence velocity) were estimated as described in 
      \citet{MFFC10}. The results are given in Table~\ref{t_respec}. In brief, temperature and gravity were measured fitting the wings of 
      strong lines known to be good indicators of these parameters. The actual values of the parameters were determined minimizing the 
      $\chi^{2}$ of the fitting with synthetic spectra drawn from the library of \citet{Coelho05}. The H$\alpha$ Balmer line was used to 
      derive the temperature \citep{Fuhrmann94}, except for the stars BT15 and BT17, which are too cool for this method. Their 
      T$_\mathrm{eff}$ was estimated from the temperature-colour relations of Alonso et al. (1999) assuming E($B-V$) = 0.02 (Schlegel, Finkbeiner 
      \& Davis 1998) and E($V-I$) = 0.03 by means of the \citet{Cardelli89} relations between reddening in different bands. Similar results 
      are obtained using the recalibration of \citet{SF11} ($A_U$ = 0.090 mag, $A_B$ = 0.075 mag, $A_V$ = 0.057 mag, $A_I$ = 0.031 mag, 
      $A_J$ = 0.015 mag, $A_K$ = 0.006 mag). Both the ($B-V$) and the ($V-I$) colours were used and the results averaged. The Ca\,{\sc i} 
      line at 6162~\AA\ \citep{Edvardsson88,Katz03}, the redder line of the Mg\,{\sc i} b triplet at 5182~\AA\ \citep{Kuijken89} and the 
      Na\,{\sc i} doublet at 5890--5893\AA\ were used for gravity measurements. The values given by the equations of \citet{Reddy03} and 
      \citet{Gratton03} were averaged to estimate the microturbulence velocity $\xi$. Finally, a total of 506 Fe\,{\sc i} lines were 
      inspected in each spectrum, with atomic data retrieved from the Vienna Atomic Line 
      Database\footnote{http://vald.astro.univie.ac.at/$\sim$vald/php/vald.php?docpage=usage.html} 
      \citep[VALD;][]{Kupka00}. For each star, 30--100 lines free of blending with other lines, in the range 5000--6500\AA, with equivalent 
      width EW=40--80~m\AA, and next to spectral regions where a clear continuum could be defined, were selected. Their EWs were measured 
      with a Gaussian fitting and they were used with the previously derived temperature, gravity and $\xi$, to estimate the metallicity, by 
      means of the local thermodynamic equilibrium (LTE) code \textsc{moog}\footnote{Freely distributed by C. Sneden, University of Texas, 
      Austin.} \citep{Sneden73}. The atmospheric models required for the calculation were obtained interpolating from the \citet{Kurucz92} 
      grid with the overshooting option switched off \citep{Castelli97}. The method was iterated when the final parameters, in particular 
      the metallicity, were noticeably different from those of the synthetic spectra used to estimate temperature and gravity. The 
      parameters were also measured with the method of the ionization/excitation equilibrium of iron lines. In this case, the temperature 
      and microturbulence velocity were fixed removing the trend of iron abundance with the excitation potential and EW, respectively, while 
      the gravity was determined by the requirements that the Fe\,{\sc i} and Fe\,{\sc ii} lines returned the same iron abundance. This 
      method was applicable only to a subset of the sample, because the low S/N of many spectra prevented us from measuring a sufficient 
      quantity of lines. Hence, we performed it only for seven stars as a safety check on the reliability of our results. A wider range of 
      line EWs (20--120~\AA) was considered in these measurements and for each star more than 100 lines were finally selected. The results 
      of this alternative method were consistent with our measurements, with only slightly higher temperature and metallicity. The mean 
      difference (in the sense our results versus ion/exc method) and rms were $\Delta$(T$_\mathrm{eff}$)=$-42\pm 120$ K, 
      $\Delta$(log(g))=0.01$\pm$0.19 dex, $\Delta$([Fe/H])=$-0.1\pm0.08$ dex. These rms should reflect the combined uncertainties of 
      the two methods, hence we obtained a rough estimate of the error (given in the last line of Table~\ref{t_respec}) of our results, 
      dividing them by $\sqrt{2}$.

      The position of each target in the temperature-gravity plane, independent of distance and reddening, was compared to Yale--Yonsei 
      isochrones \citep{Yi03} of the respective metallicity. The only free parameter was age, but its incidence on the results was 
      negligible except for subgiant stars (log(g)$\leq$4), for which the age of best isochrone fitting was also derived. The absolute 
      magnitude of each star was thus calculated, which was then used to estimate its distance. The error associated with this estimate was 
      obtained from the errors in temperature and gravity, deriving the absolute magnitude interval compatible with the 1$\sigma$ box in 
      the temperature-gravity plane. The reddening in the direction of NGC~1252 is expected to be rather negligible, E($B-V$)=0.02 
      \citep{Schlegel98,Pavani01,SF11}. For this reason, we did not estimate E($B-V$) for each target as done in \citet{MFFC10}, because 
      this quantity does not help distinguishing cluster members from field stars in this case.

      We have used the results in Table \ref{t_respec} and proper motions from the SPM4 catalogue \citep{GvA11} to calculate the Galactic 
      space velocity of the spectroscopic targets and its uncertainty (see Table \ref{kin}). The heliocentric Galactic velocity components 
      have been computed as described in \citet{JS87} for equinox 2000. Our results are referred to a right-handed coordinate system so that 
      the velocity components are positive in the directions of the Galactic Centre, U, Galactic rotation, V, and the North Galactic Pole, 
      W. We use the value of the parallax associated with the value of the distance in Table \ref{t_respec} and its error. From these values, 
      we derive the heliocentric reference frame velocity components UVW. Our spectroscopic results (see Table \ref{t_respec}) point out an 
      absence of obvious clustering in distance, radial velocity or metallicity among the brightest stars in the studied NGC~1252 area. 
      There are, however, a few of them that exhibit somewhat compatible distances (BT11, BT14, BT16, BT19 and BT21) but their radial 
      velocities are rather different and their SPM4 proper motions mutually incompatible. Another example of stars located at similar 
      distances is the pair BT12 and BT27, again with quite different radial velocities and proper motions. In principle, our high 
      resolution spectroscopy indicates that among the brightest stars in the NGC~1252 area, the majority are chemically, kinematically and 
      spatially unrelated, i.e. they appear to be part of random star field fluctuations. This scenario is typical of an asterism not of a 
      true open cluster. The only pair of stars with reasonably compatible properties is BT27 and BT28. BT27 is a subgiant that seems to be 
      located 1400$\pm$200 pc from the Sun. It is a relatively metal poor star with iron abundance of -0.28$\pm$0.07. Our two epoch 
      spectroscopy indicates that the star is single. \citet{Dias06} give a membership probability of 55 per cent for this star. BT28 is another 
      subgiant located 1300$\pm$300 pc from the Sun and relatively metal poor with iron abundance of -0.35. Our single epoch spectroscopy 
      for this star does not allow to draw any conclusions with respect to its possible binarity. Its distance, metallicity and kinematic 
      data are compatible with those of BT27 although their radial velocities are different (see Table \ref{t_respec}) but it could be a 
      binary member of the cluster, if real. \citet{Dias06} give a membership probability of 69 per cent for BT28. The pair is separated by 
      2.4 arcmin or 0.9 pc at a distance of 1350 pc. If this pair is signaling the presence of a coherent group in the area, a sequence of 
      stars with similar proper motions should be visible in the CMDs. That will be the subject of the next section.

  \section{Proper motions: candidate member selection}
\begin{figure}
     \centering
      \resizebox{\hsize}{!}{\includegraphics[]{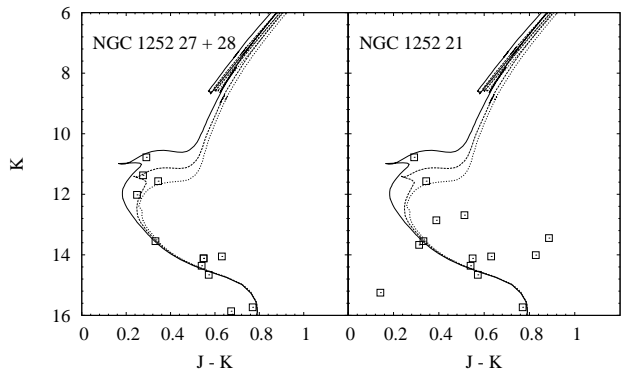}}
      \caption{2MASS near-infrared CMD for stars with proper motions within 3$\sigma$ of those of the pair NGC~1252~27 
               and NGC~1252~28 (left-hand panel) and NGC~1252~21 (right-hand panel). The dashed isochrone is for an age of 3 Gyr, [Fe/H] = -0.35 
               \citep{BMGSDRN12} and a heliocentric distance of 1.1 kpc. The continuous isochrone corresponds to an age of 2 
               Gyr and the dotted isochrone is for 4 Gyr (same iron abundance and assumed distance). 
              }
      \label{cmds}
    \end{figure}

    It is true that the brightest stars in the field of NGC~1252 do not appear to form a physical system but what if just a few of them are 
    the brightest members of a real but faint OCR? The analysis of the proper motions of stars in the area should help to answer this 
    critical question. In this work we use proper motions from the SPM4 catalogue \citep{GvA11} which contains absolute proper motions, 
    celestial coordinates, $B$, $V$ and 2MASS photometry for over 103 million stars and galaxies between the south celestial pole and 
    -20$\degr$ in declination. The catalog is roughly complete to $V$ = 17.5. We decided to use the SPM4 catalogue instead of the UCAC4 
    catalogue \citep{ZFZG11,ZFG13} because it has better coverage; for example, in a $20\times20$ arcmin$^2$ field around NGC~1252 we find 70 
    stars in UCAC4 but 116 (66 per cent more) in SPM4. On the other hand, we use SPM4 instead of the Positions and Proper Motions-eXtended Large 
    (PPMXL) catalogue \citep{RDS10} because even if it appears to be far more complete than SPM4 or UCAC4 (1148 entries for the same 
    $20\times20$ arcmin$^2$ field), the relative errors are about 50 per cent larger, rendering any analysis based on proper motions of very limited use. 
    As an example, NGC~1252~27 has UCAC4 (162-002807) proper motions ($\mu_{\alpha} \cos\delta, \mu_{\delta}) = (6.2\pm1.2, 6.3\pm1.2)$ mas 
    yr$^{-1}$, $V$ = 13.17 and $B - V$ = 0.43; its PPMXL proper motions are $\mu_{\alpha} \cos\delta$ = 16.7$\pm$11.0 mas yr$^{-1}$, 
    $\mu_{\delta}$ = 10.0$\pm$11.0 mas yr$^{-1}$, therefore somewhat compatible with UCAC4 or SPM4 data (see Table \ref{kin}) but with very 
    large errors. In this case, the UCAC4 $\mu_{\delta}$ proper motion is nearly half the value quoted by SPM4. Table \ref{kin} displays the 
    SPM4 proper motions of all the stars studied spectroscopically with $N$ being the number of stars in the SPM4 catalogue sharing the 
    proper motion -- within 3$\sigma$ of the values quoted -- of a given star in the table. Let us assume that one of these stars is a 
    member of NGC~1252. The only star with a relatively large cohort of candidate comoving stars is NGC~1252~21, a star of almost solar 
    metallicity located at a distance of 360$\pm$40 pc from the Sun. The proper motions of this star match well those attributed to NGC~1252 
    in \citet{Dias06} and \citet{ZPBML12}. Unfortunately, the CMD (see Fig. \ref{cmds}, right-hand panel) suggests that the 
    candidate group is made of stars located at different distances (perhaps also of different metallicities): some as close as NGC~1252~21 
    but others located as far away as 1~kpc. Although the individual number of candidate comoving stars for NGC~1252~27 and NGC~1252~28 is 
    smaller, their combined figure is close to the value of $N$ found for NGC~1252~21. If, as some studies suggest (see Section 2), there is 
    a real open cluster located at a heliocentric distance close to 1~kpc in the area of sky commonly associated to NGC~1252 and we assume 
    that the stars NGC~1252~27 and NGC~1252~28 are among its brightest members, the 2MASS CMD in Fig. \ref{cmds}, left-hand 
    panel, supports that conclusion. It also shows that some of the stars with proper motions compatible with those of NGC~1252~21 are also 
    compatible with those of the pair NGC~1252~27 and NGC~1252~28 as their 3$\sigma$ spreads overlap, even if their distances are rather 
    different. For example, the second brightest star (the brightest is NGC~1252~28) in Fig. \ref{cmds}, left-hand panel, is TYC 8498-853-1 which 
    is an F5V star located 533 pc from the Sun \citep{PD10}; this star has a 75 per cent membership probability in \citet{Dias06}. The third 
    brightest star in the same diagram is NGC~1252~21 itself. Both TYC 8498-853-1 and NGC~1252~21 are not part of NGC~1252, as described 
    here. This is to be expected within the context of foreground stellar contamination, as already pointed out in Section 2. The main 
    source of this foreground contamination could be the Hyades stream as the velocity components of NGC~1252~21 in Table \ref{kin} are very 
    close to the range quoted by \citet{Pom11}. This somewhat coherent structure is not a cluster but the result of resonances. 
%
%
%
%
%
%
    \begin{figure}
     \centering
      \resizebox{\hsize}{!}{\includegraphics[]{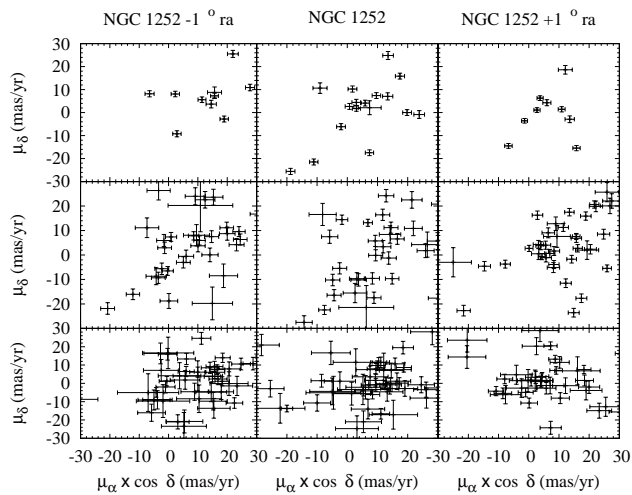}}
      \caption{Proper motion analysis: proper motion components with 1$\sigma$ error bars in a $20\times20$ arcmin$^2$ field around 
               NGC~1252 and two equivalent fields centred at the position of NGC~1252 $\pm1\degr$ in right ascension.
               {\it Top panel:} stars brighter than $K = 12$.
               {\it Middle panel:} stars having 12 $ \leq K < 14$. 
               {\it Bottom panel:} stars having 14 $ \leq K < 16$. 
              }
      \label{alla}
    \end{figure}
%
%
%
%
    \begin{figure}
     \centering
      \resizebox{\hsize}{!}{\includegraphics[]{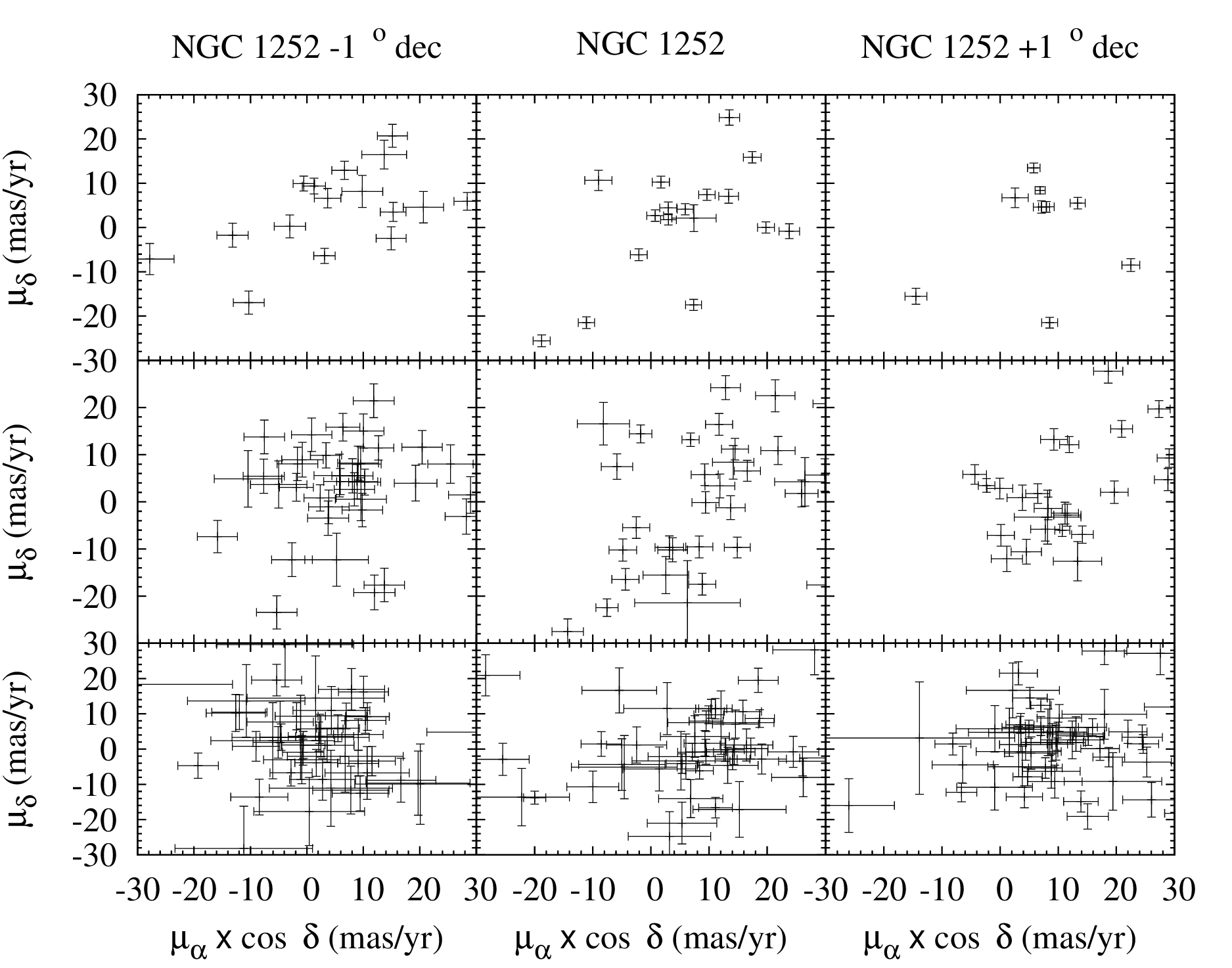}}
      \caption{Same as Fig. \ref{alla} but for a $20\times20$ arcmin$^2$ field around NGC~1252 and two equivalent fields centred at the position 
               of NGC~1252 $\pm1\degr$ in declination.
              }
      \label{allb}
    \end{figure}
%
%
%
%
    \begin{figure}
     \centering
      \resizebox{\hsize}{!}{\includegraphics[]{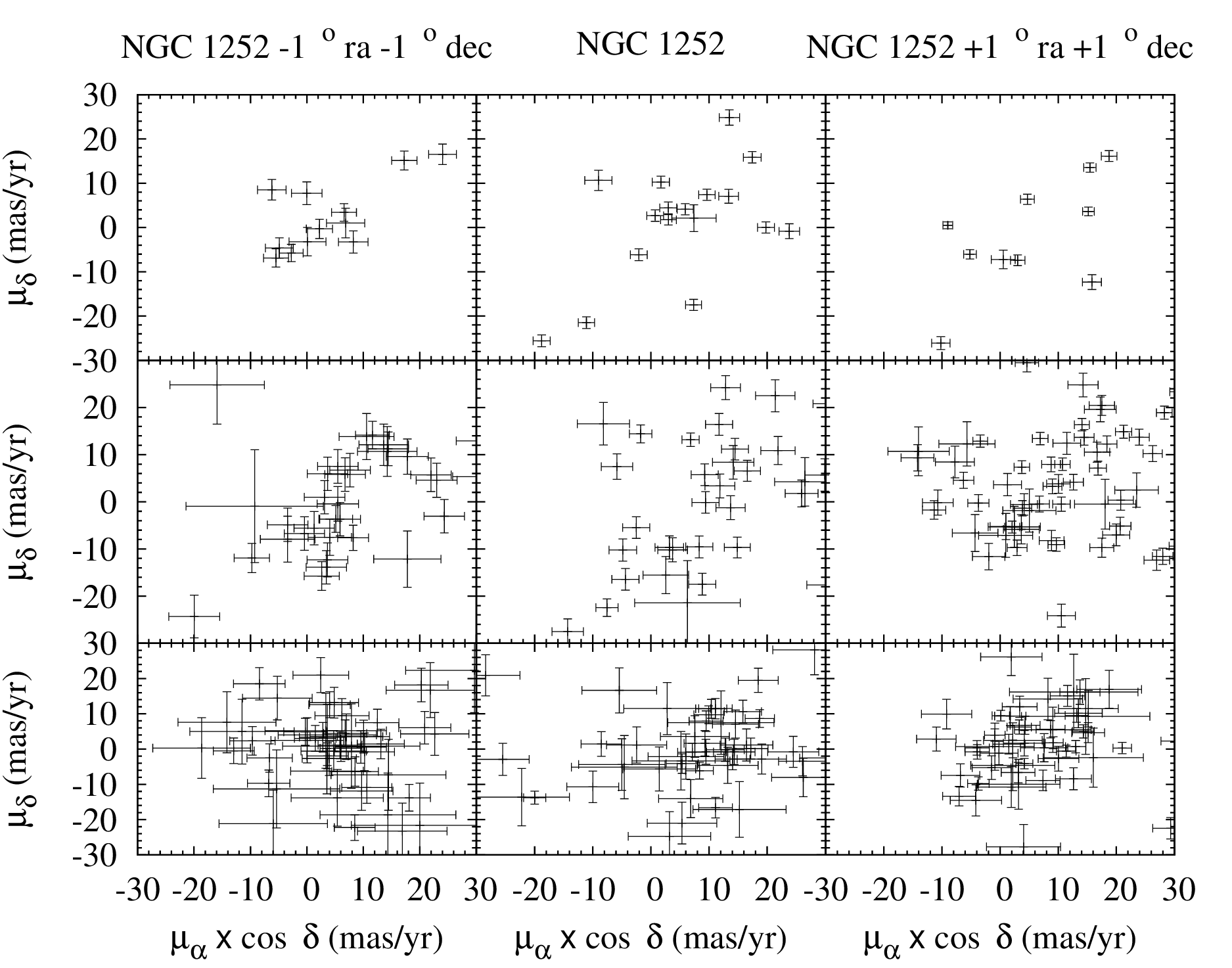}}
      \caption{Same as Fig. \ref{alla} but for a $20\times20$ arcmin$^2$ field around NGC~1252 and two equivalent fields centred at the position 
               of NGC~1252 $-1\degr$ in right ascension, $-1\degr$ in declination and $+1\degr$ in right ascension, $+1\degr$ in declination.
              }
      \label{allc}
    \end{figure}
%
%
%
%
    \begin{figure}
     \centering
      \resizebox{\hsize}{!}{\includegraphics[]{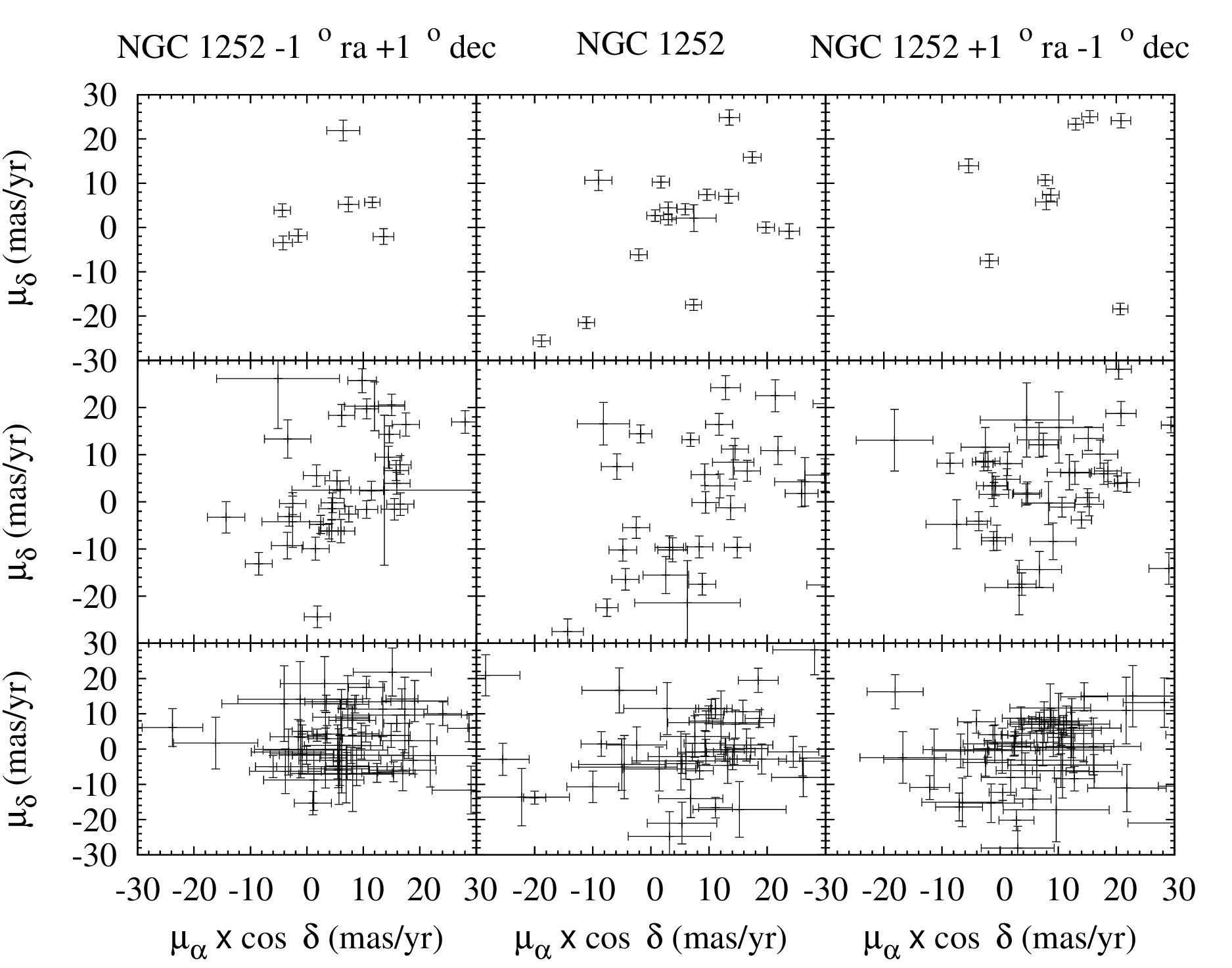}}
      \caption{Same as Fig. \ref{alla} but for a $20\times20$ arcmin$^2$ field around NGC~1252 and two equivalent fields centred at the position
               of NGC~1252 $-1\degr$ in right ascension, $+1\degr$ in declination and $+1\degr$ in right ascension, $-1\degr$ in declination. 
              }
      \label{alld}
    \end{figure}
%
%

    Under normal circumstances, overdensities in proper motion diagrams signal the presence of kinematically coherent structures. The usual 
    approach followed in \citet{CFVMFBS07} consists in extracting proper motion components in a $20\times20$ arcmin$^2$ field around the object 
    and search for statistically significant clumps. Alternatively, we can also compare with adjacent star fields looking for peculiar 
    features that are only present in one area. Figs \ref{alla}--\ref{alld} show three vector-point diagrams as a function of the $K$ magnitude 
    for a $20\times20$ arcmin$^2$ field around NGC~1252 and eight equivalent neighbouring star fields. The top panel is restricted to stars having 
    $K<$ 12, the middle panel to stars in the range $12 \leq K < 14$ and the bottom panel to stars having $14 \leq K < 16$. Stars in the 
    top panel for the NGC~1252 area appear to clump around $\mu_{\alpha} \cos\delta$ = 4 mas yr$^{-1}$, $\mu_{\delta}$ = 3 mas yr$^{-1}$ but 
    that is not observed at fainter magnitudes (see Fig. \ref{alla}, three central panels). However, for stars with $12 \leq K < 14$, a 
    group is visible centred at $\mu_{\alpha} \cos\delta$ = 12 mas yr$^{-1}$, $\mu_{\delta}$ = 6 mas yr$^{-1}$ which are close to the values 
    cited by \citet{Dias06} and \citet{ZPBML12} and a more extended structure is observed at fainter magnitudes. These results, together 
    with the analysis of Fig. \ref{cmds} and Table \ref{kin}, indicate that the clump found is the result of two contributions, the $\sim$1 
    kpc distant NGC~1252 and a significantly closer population that we interpret as part of the Hyades stream. Given the small number of 
    stars involved and the tight mixture of unrelated (in terms of heliocentric distance and metallicity) populations in proper motion space, 
    a simple statistical analysis looking for overdensities in Figs \ref{alla}--\ref{alld} does not draw any meaningful conclusions.    

\begin{figure*}
     \centering
      \resizebox{\hsize}{!}{\includegraphics[]{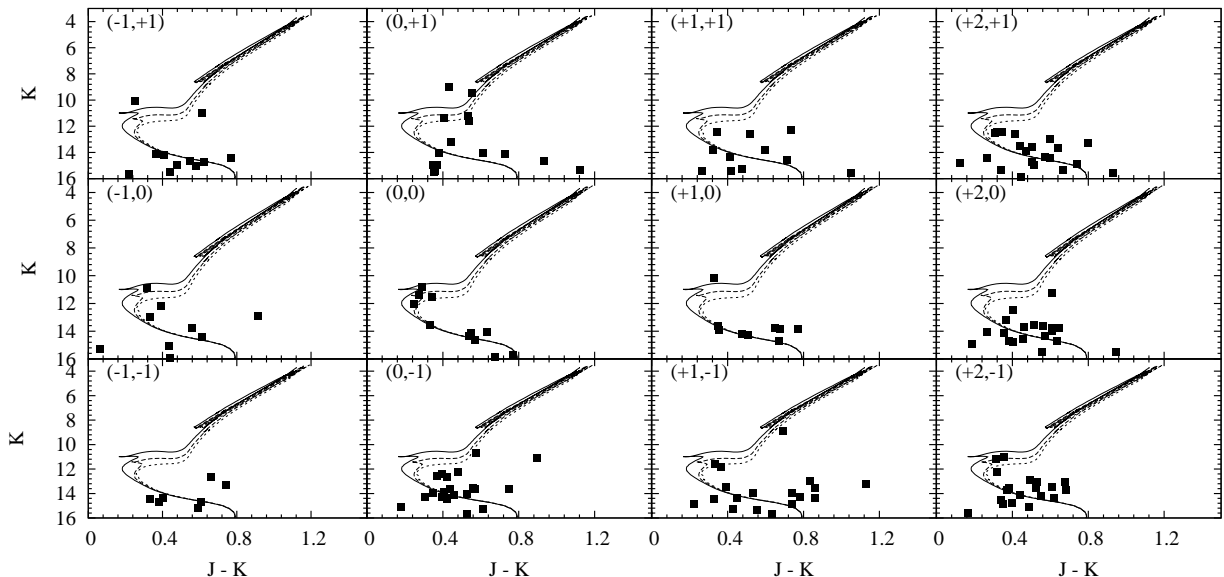}}
      \caption{2MASS near-infrared CMDs for stars with proper motions in SPM4 within 3$\sigma$ of those of the pair 
               NGC~1252~27 and NGC~1252~28 at various locations around NGC~1252. Each panel corresponds to a $20\times20$ arcmin$^2$ patch of sky 
               centred at the position of NGC~1252 and shifted in right ascension and declination by the amounts indicated in parentheses
               (top left-hand corner). For example, (+1, -1) means that the squared patch of sky is centred at the location of NGC~1252 
               $+1\degr$ in right ascension and $-1\degr$ in declination.   
              }
      \label{cmdsaround}
    \end{figure*}

    Out of all the bright stars studied spectroscopically in this work, only two appear to be kinematically consistent with membership on a 
    possible NGC~1252 stellar group but their radial velocities are different and we cannot confirm the single nature of NGC~1252~28 as we 
    have single epoch spectroscopy. There are 11 stars (out of a total of 116 in SPM4, $\sim$10 per cent) with proper motions within 3$\sigma$ 
    of those of NGC~1252~27 and NGC~1252~28 in our $20\times20$ arcmin$^2$ field centred on the SIMBAD's coordinates of NGC~1252. The 2MASS 
    near-infrared CMD for these suspected NGC~1252 members is seen in Fig. \ref{cmds}, left-hand panel. The best Padova 
    theoretical isochrone\footnote{http://stev.oapd.inaf.it/cgi-bin/cmd} fitted corresponds to an age of 3 Gyr and abundance [Fe/H] = -0.35
    \citep{BMGSDRN12}, assuming a heliocentric distance of nearly 1 kpc. Isochrones within an age range of 2--4 Gyr and iron abundance range 
    of -0.20 to -0.50 are also compatible; a heliocentric distance to NGC~1252 $\gg$ 1 kpc appears to be highly unlikely. But, how 
    significant is the group found? If we study the CMDs for stars with proper motions within 3$\sigma$ of those of 
    the pair NGC~1252~27 and NGC~1252~28 at various locations in the neighbourhood of NGC~1252, we can find out if the sequence in Fig. 
    \ref{cmds}, left-hand panel, is peculiar to the area observed or shared across a wider region. Fig. \ref{cmdsaround} suggests that members 
    of a kinematically coherent but extended structure, the associated moving group, occupy an area approximately located between right 
    ascension $3^{\rm h}~10^{\rm m}$ and $3^{\rm h}~15^{\rm m}$ with declination between $-57\degr 34\arcmin$ and -57\degr 58\arcmin; it 
    seems to extend east of NGC~1252. In summary, the OCR that may be present in the area of NGC~1252 is relatively close ($\sim$1 kpc), of 
    subsolar metallicity and probably old. In Tables \ref{FINAL_T}, we have compiled relevant data for the 11 suspected NGC~1252 members 
    plotted in Fig. \ref{cmds}, left-hand panel, and two other candidates (see below). Our average proper motions for NGC~1252 are $\mu_{\alpha} 
    \cos\delta$ = 9.2$\pm$3.0 mas yr$^{-1}$, $\mu_{\delta}$ = 8.8$\pm$2.8 mas yr$^{-1}$. These values are consistent with those in 
    \citet{ZPBML12}. The cluster centre is found at $\alpha=03^{\rm h}~10^{\rm m}~47^{\rm s}$, $\delta=-57\degr 45\arcmin 18\arcsec$. 
    
%
%
%
%

  \section{NGC 1252: a high Galactic altitude open cluster remnant}
\begin{figure*}
     \centering
      \includegraphics[width=\textwidth]{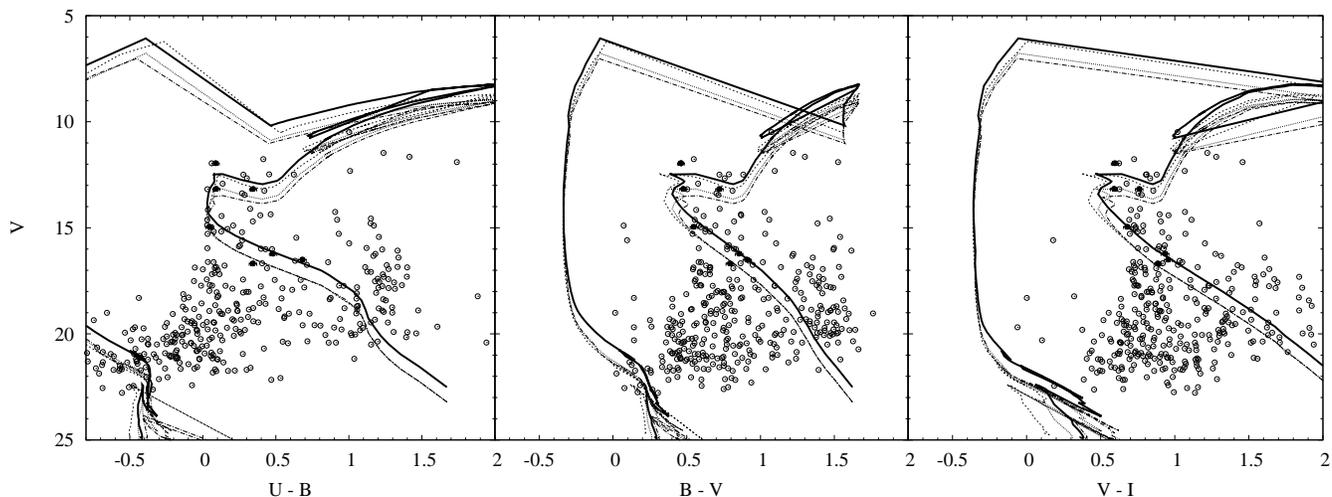}
      \caption{Optical CMDs for the photometry presented in Section 3.1 and the candidate members discussed in Section 
               4. In the three panels, the isochrones correspond to a metallicity of [Fe/H] = -0.35 and ages 2, 3 and 4 Gyr from models by
               \citet{GBBC00} which include the post-giant evolution and the locus of the white dwarf cooling sequence. Traces of a possible
               white dwarf cooling sequence can be seen in the first two diagrams. The seven stars in common with those in the list of 11 
               suspected NGC~1252 members based on proper motions appear as solid points. An heliocentric distance of 1350 pc has been 
               assumed but the solid, thick line is the isochrone of age 3 Gyr at a distance of 1 kpc.
              }
      \label{ourscmd}
    \end{figure*}

    In Fig. \ref{ourscmd}, we present several optical CMDs obtained from the photometry presented in Section 3.1 and 
    using the values of the OCR parameters constrained by the proper motion analysis carried out in Section 4. The stars with proper motions 
    within 3$\sigma$ of those of NGC~1252~27 and NGC~1252~28 are plotted as filled black squares. Our photometry is plotted as empty 
    circles. Typical of OCRs, there is a sparse main sequence and the turnoff point is not well defined but it could be near $V$ = 12. 
    Hardly any trace of the red giant branch or red clump is observed. In principle, they could be absent because the associated stars were 
    saturated on our images. No blue stragglers are observed but this could also be the result of image saturation. The large number of 
    objects under the proposed main sequence is made of faint field stars, probably part of the distant halo like NGC~1252~17 (see Appendix
    A). The colour of the bluest of these stars is redder than the proposed location of the turnoff point of NGC~1252 so they could be 
    older than the OCR but they could also have higher (perhaps, solar) metallicity. The isochrones from \citet{GBBC00} correspond to a 
    metallicity of [Fe/H] = -0.35 and ages 2, 3 and 4 Gyr; they include the post-giant evolution and the locus of the white dwarf cooling 
    sequence. An age of 3 Gyr is favoured but differences are not very significant and they are probably well within the observational 
    errors. An interesting feature that is glimpsed in the first two diagrams, is the presence of a population of faint blue stars that 
    appear in a sequence near $V$ = 22. They could be faint unresolved galaxies but objects there could also be candidates to signal the 
    locus of the NGC~1252 white dwarf cooling sequence. They are well below the completeness limit, however. $N$-body simulations suggest 
    that up to 10 per cent of members of old OCRs could be white dwarfs if the population of the original cluster was larger than a few thousand 
    stars \citep[see fig. 3 in][]{dlFM98}. Deeper photometry is needed to confirm this tentative analysis. 
%
%
%
%

    In order to find additional candidates for membership in NGC~1252 and confirm the lack of red giant branch or red clump, we have 
    searched SPM4 for objects with proper motions within 3$\sigma$ of those of NGC~1252~27 and NGC~1252~28 and distance to the centre of 
    the cluster $<$ 12 arcmin (over 3 pc at the assumed distance for the cluster). This search gives 13 objects (including the ones already 
    discussed above) that are included in Table \ref{FINAL_T}. In order to find out which objects are the most relevant members of the 
    cluster we have to estimate a membership probability. A robust choice to estimate this probability from proper motions is a bivariate 
    Gaussian distribution. The probability of object $i$ being a member of a kinematic group using $\mu_{\alpha}^{*} = \mu_{\alpha} 
    \cos\delta$ and $\mu_{\delta}$ is given by:
    \begin{equation}
       P^{k}_i = {\Large e}^{-\frac{1}{2}\left[
                   \left(\frac{\mu_{\alpha i}^{*} - <\mu_{\alpha}^{*}>}{\sigma_{\mu_{\alpha}^{*}}}\right)^{2} +
                     \left(\frac{\mu_{\delta i} - <\mu_{\delta}>}{\sigma_{\mu_{\delta}}}\right)^{2}
                   \right]} \,,
                   \label{kinpro}
    \end{equation}
    where $<\mu_{\alpha}^{*}>$ (9.2 mas yr$^{-1}$), $<\mu_{\delta}>$ (8.8 mas yr$^{-1}$) are the average values for the cluster,
    $\sigma_{\mu_{\alpha}^{*}}$ (3.0 mas yr$^{-1}$), $\sigma_{\mu_{\delta}}$ (2.8 mas yr$^{-1}$) are the associated standard deviations and 
    $\mu_{\alpha i}^{*}$, $\mu_{\delta i}$ are the proper motions of the $i$th object. The computed values are given in Table \ref{FINAL_T}, 
    last column; 3 stars have membership probability higher than 80 per cent. Given the uncertainty in the possible value of the radial velocity of 
    the cluster (we do not have three or more stars with similar radial velocities), we do not compute the motion or Galactic space velocity 
    of NGC~1252. In addition, we do not estimate its present-day mass function or initial mass function (IMF) as we do not have enough 
    spectroscopically confirmed members to obtain a meaningful result. For the same reason, we do not try to provide a value for the cluster 
    effective radius, the tidal radius or the concentration parameter. We can however conclude that the degree of concentration of this OCR 
    is rather low as expected of an old and relaxed dynamical system. As for the original population of the cluster, its current population 
    (see Fig. \ref{ourscmd}) could be $\sim$30 members and for such membership after 3 Gyr, the results of numerical simulations 
    \citep{dlFM98} suggest that the initial membership could be as high as 10$^{4}$ stars, much larger than our estimate for NGC~1901 in 
    \citet{CFVMFBS07}. This estimate puts the parent open cluster of the present NGC~1252 OCR among the most massive open clusters currently 
    observable in the Milky Way. Further discussion on the possible origin of this unusual object is presented below. 

%
%
\begin{table*}
 \fontsize{8} {10pt}\selectfont
 \tabcolsep 0.10truecm
 \caption{Candidates for membership in NGC~1252 based on SPM4 proper motions. The distance from the customary centre of the 
          cluster in arcminutes is represented by $r$. The entries are sorted by $r$.} 
 \resizebox{\linewidth}{0.15\linewidth}{
 \begin{tabular}{cccccccccccccccc}
 \hline
 \multicolumn{1}{c}{ID}                         &
 \multicolumn{2}{c}{Designation}                &
 \multicolumn{1}{c}{$r$}                        &
 \multicolumn{1}{c}{$\alpha$}                   &
 \multicolumn{1}{c}{$\delta$}                   &
 \multicolumn{1}{c}{$\mu_{\alpha} \cos\delta$}  &
 \multicolumn{1}{c}{$\mu_{\delta}$}             &
 \multicolumn{1}{c}{$U$}                        &
 \multicolumn{1}{c}{$B$}                        &
 \multicolumn{1}{c}{$V$}                        &
 \multicolumn{1}{c}{$I$}                        &
 \multicolumn{1}{c}{$J$}                        &
 \multicolumn{1}{c}{$H$}                        &
 \multicolumn{1}{c}{$K$}                        &
 \multicolumn{1}{c}{$P^{k}_i$}                  \\
 \hline
     & SIMBAD & SPM4 & (arcmin) & $(^{\rm h}:^{\rm m}:^{\rm s})$ & ($\degr$:$\arcmin$:$\arcsec$) & (mas yr$^{-1}$) & (mas yr$^{-1}$) & (mag) & (mag) & (mag) & (mag) & (mag) & (mag) & (mag) & (per cent)  \\
 \hline
  122 & NGC 1252 28    & 1230015448 &  1.93 & 03:10:38.5 & -57:47:19 &  9.6$\pm$1.4 &  7.4$\pm$1.3 & 12.50$\pm$0.04 & 12.41$\pm$0.03 & 11.95$\pm$0.02 
      & 11.35$\pm$0.03 & 11.07$\pm$0.04 & 10.87$\pm$0.04 & 10.78$\pm$0.03 & 87.2 \\
  176 & NGC 1252 27    & 1230015460 &  2.05 & 03:10:56.6 & -57:47:47 &  6.8$\pm$1.5 & 13.2$\pm$1.4 & 13.73$\pm$0.04 & 13.64$\pm$0.03 & 13.16$\pm$0.02 
      & 12.57$\pm$0.03 & 12.27$\pm$0.03 & 12.00$\pm$0.03 & 12.02$\pm$0.03 & 22.5 \\
  227 & -              & 1230083297 &  3.02 & 03:11:11.2 & -57:46:34 & 12.1$\pm$9.1 &  7.5$\pm$9.0 & 18.73$\pm$0.04 & 18.68$\pm$0.03 & 17.97$\pm$0.02
      & 17.07$\pm$0.03 & 16.50$\pm$0.14 & 15.91$\pm$0.17 & 15.73$\pm$0.24 & 56.0 \\
  236 & -              & 1230082026 &  4.36 & 03:11:16.2 & -57:48:25 &  2.8$\pm$7.5 & 11.5$\pm$7.4 & 18.00$\pm$0.04 & 18.06$\pm$0.03 & 17.51$\pm$0.02
      & 16.79$\pm$0.03 & 16.54$\pm$0.13 & 16.15$\pm$0.22 & 15.86$\pm$0.25 & 06.6 \\
  192 & -              & 1230084516 &  4.76 & 03:10:59.9 & -57:41:28 & 11.1$\pm$2.9 & 11.5$\pm$2.9 & 18.08$\pm$0.04 & 17.40$\pm$0.03 & 16.50$\pm$0.02
      & 15.54$\pm$0.03 & 14.90$\pm$0.04 & 14.42$\pm$0.04 & 14.36$\pm$0.07 & 52.2 \\
   22 & -              & 1230083252 &  6.72 & 03:09:58.9 & -57:45:17 &  9.5$\pm$2.9 &  7.9$\pm$2.9 & 17.81$\pm$0.04 & 17.47$\pm$0.03 & 16.68$\pm$0.02
      & 15.79$\pm$0.03 & 15.23$\pm$0.06 & 14.80$\pm$0.06 & 14.66$\pm$0.10 & 94.2 \\
   99 & -              & 1230015445 &  7.62 & 03:10:32.1 & -57:38:43 &  9.2$\pm$2.3 &  5.8$\pm$2.3 & 15.54$\pm$0.04 & 15.50$\pm$0.03 & 14.95$\pm$0.02 
      & 14.27$\pm$0.03 & 13.88$\pm$0.03 & 13.58$\pm$0.03 & 13.55$\pm$0.04 & 55.5 \\
  279 & NGC 1252 21    & 1230015481 &  8.43 & 03:11:36.0 & -57:40:23 & 13.4$\pm$1.7 &  7.0$\pm$1.6 & 14.23$\pm$0.04 & 13.89$\pm$0.03 & 13.17$\pm$0.02
      & 12.41$\pm$0.03 & 11.91$\pm$0.02 & 11.61$\pm$0.03 & 11.57$\pm$0.02 & 30.3 \\
   -  & TYC 8498-853-1 & 1230000655 &  9.55 & 03:09:44.9 & -57:50:16 &  5.9$\pm$1.4 &  4.1$\pm$1.3 &  -             & -              & -
      & -              & 11.65$\pm$0.02 & 11.42$\pm$0.03 & 11.37$\pm$0.02 & 13.9 \\
  340 & -              & 1230015496 & 10.40 & 03:12:04.1 & -57:43:13 &  9.0$\pm$2.3 & 14.5$\pm$2.3 & 15.63$\pm$0.04 & 15.57$\pm$0.03 & 14.97$\pm$0.02 
      & 14.24$\pm$0.04 & 13.65$\pm$0.03 & 13.29$\pm$0.03 & 13.30$\pm$0.05 & 13.3 \\
   -  & -              & 1230081981 & 10.86 & 03:09:37.5 & -57:51:14 &  9.9$\pm$2.5 &  9.7$\pm$2.5 &  -             & -              & -
      & -              & 14.66$\pm$0.04 & 14.23$\pm$0.04 & 14.12$\pm$0.06 & 91.8 \\
   -  & -              & 1230015446 & 10.92 & 03:10:35.3 & -57:56:45 & 10.6$\pm$2.7 &  4.0$\pm$2.7 &  -             & -              & -
      & -              & 14.47$\pm$0.03 & 14.21$\pm$0.04 & 14.05$\pm$0.06 & 21.0 \\
  335 & -              & 1230084546 & 11.06 & 03:12:02.5 & -57:40:53 & 10.4$\pm$2.7 & 11.4$\pm$2.7 & 17.55$\pm$0.04 & 17.07$\pm$0.03 & 16.22 $\pm$0.02
      & 15.28$\pm$0.03 & 14.68$\pm$0.04 & 14.17$\pm$0.05 & 14.05$\pm$0.07 & 60.5 \\
 \hline
 \end{tabular}
 }
 \label{FINAL_T}
\end{table*}
%
%

    How do our results compare with those obtained in past studies of this area? We agree with \citet{Bouchet83} that there is an actual 
    open cluster in that region of the sky but in contrast with our own findings, their conclusions draw a picture of a younger and closer 
    object. That picture is not supported by our current results. We disagree with \citet{Eggen84} as he concluded that NGC~1252 was 
    probably not a cluster (more on this later). We cannot confirm that NGC~1252 has any blue stragglers as pointed out by \citet{Ahumada95}, 
    \citet{XD05} and \citet{Ahumada07}: HD~20286 is a foreground star (see Appendix A). We do not agree with the conclusions obtained by 
    \citet{Baumgardt98} and \citet*{Baumgardt00}: our results indicate that NGC~1252 is a real open cluster, not an asterism. We coincide 
    with the conclusions drawn by \citet{Pavani01} and \citet{BSDDOP01} regarding the presence of an OCR in the area and its age but their 
    distance is almost half the value found in our study. As a consequence, their object is closer to the Galactic plane. This smaller value 
    of the distance could be the result of the relatively large differences in $V$ between our photometry and theirs as pointed out in 
    Section 3.1.4: their stars brighter than $V$ = 13 are up to 0.3 mag brighter than ours. It could also be possible that \citet{Pavani01} 
    and \citet{BSDDOP01} identified the population that we consider part of the Hyades stream as NGC~1252. These stars have a significant 
    distance spread but they are closer than 600 pc and they are present in all the extended area around the cluster (see Fig. 
    \ref{cmdsaround}). Similar comments can be made about \citet{LB03}, their proper motions are also slightly smaller than the values
    obtained here. In contrast, our proper motion results agree well with those in \citet{Dias06} and \citet{ZPBML12}. Nonetheless, we 
    believe that these authors are mixing stars located at different distances, including many candidate members that are, in fact, part of 
    the foreground population. As for the value of the diameter quoted in \citet{vdBergh06}, our results suggest a larger value as the 
    object exhibits no concentration but, as pointed out above, more data are needed to provide a reliable figure. Finally, \citet{PB07} use 
    2MASS photometry and UCAC2 proper motions. This may explain the shorter distance obtained by these authors in comparison with our result. 
    In general, the somewhat shorter distance quoted in some papers could also be the result of assuming that the cluster has solar 
    metallicity which appears not to be the case.
    
    Regarding the origin of NGC~1252, \citet{vdBergh06} pointed out that open clusters with ages $>$ 1 Gyr appear to form a singular 
    structure that he termed a `cluster thick disc'. Van den Bergh considers that part of the open cluster thick disc consists of objects 
    that were probably captured gravitationally by the main body of the Galaxy. Similar views appear in \citet{GKM12}. This does not 
    necessarily imply that a fully formed star cluster was captured; our Galaxy harbors a population of high Galactic altitude gas clouds 
    which are, in some cases, similar to those found in the disc. Although the disc hosts nearly 95 per cent of all known young open clusters and 
    even a larger fraction of GMCs, a small but significant percentage of young open clusters are found far from the thin disc \citep{RC08}. 
    Within the disc, star formation is mainly driven by the large-scale shock induced by a spiral arm although supernovae also play a 
    non-negligible role. However in absence of spiral shocks, what mechanism could form star clusters at high Galactic altitude? This was 
    one of the objections pointed out originally by \citet{Eggen84} to conclude that NGC~1252 was probably not a cluster because it was 
    located too far from the disc. Contrary to this old, restrictive view, star formation at high Galactic altitude is now well documented 
    observationally \citep[see][for details]{RC08} and several mechanisms capable of explaining its existence have been suggested: 
    supernova-triggered star formation \citep[e.g.][]{WBLH77}, off-plane gass ejection \citep{MAFK99} and tidal encounters \citep{RC08}. 
    These mechanisms can operate concurrently. Star formation far from the Galactic disc is not unusual and \citet{RC08} already concluded 
    that nearly 13 per cent of known open clusters are found at $|z| \geq$ 200 pc, where $z = d \sin b$, $d$ and $b$ are the heliocentric distance 
    and Galactic latitude, respectively, of the object. The 2013 January version \citep[v3.3;][]{Dias13} of NCOVOCC includes 2174 open 
    clusters; out of them, 1629 have both distance and age. Within this smaller sample, we find 293 (18 per cent) clusters of all ages with 
    separation from the disc $>$ 200 pc. Among young clusters (age $<$ 100 Myr) the fraction of high-altitude objects is 4.7 per cent (24 out of 
    505 clusters). This fraction increases to 50.6 per cent (158 out of 312 clusters) for old clusters (age $>$ 1 Gyr). As expected, moving in an 
    inclined orbit increases the survival opportunities of a star cluster. These numbers clearly vindicate the conclusions of 
    \citet{vdBergh06} pointed out above and also single out NGC~1252 among known OCRs. With a current distance below the Galactic disc of 
    almost 900 pc, NGC~1252 is the first object that can be truly considered a high Galactic altitude OCR. The orbit of the parent GMC of 
    NGC~1252 may have been unusually inclined perhaps as a result of the Galactic warp or a tidal interaction with another massive object. 
    An alternative scenario is in situ formation, maybe as a result of a tidally induced star formation event at high Galactic altitude 
    following the tidal encounter paradigm outlined by \citet{RC08}: a close encounter between a high Galactic altitude gas cloud and a 
    globular cluster.

  \section{Conclusions}
    In this paper, we have reexamined the open cluster remnant nature of NGC~1252 in Horologium. This controversial object had been 
    classified both as asterism \citep[e.g.][]{Baumgardt98} and OCR \citep[e.g.][]{Pavani01}. The main objective of this research was to 
    shed some new light on this controversy and we believe that we have accomplished this objective as well as clarified and discussed some 
    other important aspects related to this area of the sky. Our main conclusions can be summarized as follows: 
    \begin{itemize}
      \item All the available evidence shows that there is a very sparse, faint open cluster or OCR 
            in the area of sky commonly associated to NGC~1252.
      \item This poorly populated stellar group is located at about 1 kpc from the Sun and it is 3 
            Gyr old.
      \item Spectroscopy of two candidate members indicates that the group has subsolar metallicity.
      \item We identify about a dozen candidate members but further observations are required in 
            order to confirm membership in terms of photometry, spectroscopy and astrometry. In 
            particular, deeper photometry and multi-epoch spectroscopy of the brightest candidate 
            members are needed.
      \item In light of numerical simulations \citep{dlFM98}, what is observed now at the location 
            of NGC~1252, the OCR, is compatible with an original open cluster made of perhaps as 
            many as 10$^{4}$ stars.
      \item With a current distance below the Galactic disc of almost 900 pc, NGC~1252 is the first 
            open cluster that can be truly considered a high Galactic altitude OCR; an unusual 
            object that can hint at a star formation episode induced on a high Galactic altitude gas 
            cloud by a non-standard mechanism, perhaps a passing globular cluster \citep{RC08}.
      \item The carbon star TW~Horologii, associated by some authors with the open cluster NGC~1252, 
            is clearly a foreground object. Its distance and kinematic properties are incompatible 
            with membership in NGC~1252.
      \item The blue straggler candidate HD~20286, identified by multiple authors as a true member 
            of NGC~1252, is a foreground object. Its status as blue straggler should be revised.
      \item NGC~1252~21 that has been commonly associated with the centre of NGC~1252 appears to be 
            a foreground star, unrelated to the cluster.
      \item NGC~1252~1 (HD 20059) appears to be a relatively close star, the primary of a wide 
            binary system.
      \item NGC~1252~17 is a background Population II star moving in a retrograde orbit at a very 
            high speed, 403 km s$^{-1}$, but probably still bound to the Milky Way 
            \citep[see][]{Smith07}. It could be a cannonball star (Meylan, Dubath \& Mayor 1991).
    \end{itemize} 
    As a metal poor open cluster of Gyr age, this object is of interest to studies of stellar evolution and star formation. The study of 
    OCRs is always a challenging but rewarding endeavour; however and in the case of an object like NGC~1252, the high Galactic latitude 
    (and, in this case, also altitude) of the object and significant foreground contamination complicate matters further. As pointed out 
    above, deeper photometry and additional multi-epoch spectroscopy are needed in order to improve the values of the parameters of this 
    object now that its true nature has been better stablished.  

  \section*{Acknowledgements}
    The authors thank the referee for his/her constructive reports and helpful suggestions regarding the presentation of this paper. 
    RdlFM and CdlFM acknowledge partial support by the Spanish `Comunidad de Madrid' under grant CAM S2009/ESP-1496. EC acknowledges support by the 
    Fondo Nacional de Investigaci\'on Cient\'ifica y Tecnol\'ogica (proyecto No. 1110100 Fondecyt) and the Chilean Centro de Excelencia en 
    Astrof\'isica y Tecnolog\'ias Afines (PFB 06). In preparation of this paper, we made use of the NASA Astrophysics Data System and the 
    ASTRO-PH e-print server. This work also has made extensive use of the SIMBAD data base and the VizieR catalogue access tool, both 
    operated at the CDS, Strasbourg, France. This publication makes use of data products from the Two Micron All Sky Survey, which is a 
    joint project of the University of Massachusetts and the Infrared Processing and Analysis Center/California Institute of Technology, 
    funded by the National Aeronautics and Space Administration and the National Science Foundation. 

  \bibliographystyle{mn2e}

  \appendix
  \section{A few objects in the NGC 1252 field}
    The area around NGC~1252 includes two objects that have been associated to NGC~1252 by some authors. Here we compile available 
    data to confirm or disprove their connection to NGC~1252. We also discuss a few other objects with unusual properties.

    \subsection{TW Horologii}
      TW~Horologii (TW Hor; =HD~20234=HR~977) is a naked-eye variable orange star located nearly 30 arcmin north-east of NGC~1252 (see Fig. 
      \ref{ngc1252field} for their relative positions). It is one of the sky's brightest carbon stars normally shining at $V$ magnitude 5.79 mag
      but displaying small variations of 0.6 mag with a period of 157 d \citep{Bouchet84}. \citet{Bouchet83} first suggested that TW~Hor 
      was a probable member of NGC~1252 but shortly after, \citet{Eggen84} contested their result by arguing that NGC~1252 was not a true 
      open cluster but just an asterism. For \citet{Eggen84}, TW~Hor is a member of the Hyades supercluster. TW~Hor is bright enough to be 
      studied by the Hipparcos mission. The new Hipparcos reduction \citep{vanLeeuwen07} gives a distance of 322$^{+43}_{-34}$ pc to TW~Hor 
      that is incompatible with our distance determination for the cluster. Proper motions, ($\mu_{\alpha} \cos\delta, \mu_{\delta}) = 
      (18.44\pm0.31, 13.20\pm0.35)$ mas yr$^{-1}$, also from the new Hipparcos reduction, are inconsistent with membership too. The Extended 
      Hipparcos Compilation (XHIP) which is based on the new Hipparcos reduction gives a distance of 316$\pm$38 pc with ($\mu_{\alpha} 
      \cos\delta, \mu_{\delta}) = (18.49\pm0.29, 13.26\pm0.31)$ mas yr$^{-1}$, that translates into a transverse velocity of 34.1 km s$^{-1}$ 
      \citep{Anderson12}. Its radial velocity is 14.3$\pm$2.9 km s$^{-1}$ \citep{Gontcharov06} which is incompatible (within the error 
      limits) with that of NGC~1252~27 or NGC~1252~28 (see Table~\ref{t_respec}). All this body of observational evidence indicates that 
      TW~Hor is not part of NGC~1252. On the other hand, both its radial velocity and distance are compatible with those of NGC~1252~19 
      in Table~\ref{t_respec} but their proper motions are fully incompatible. TW~Hor's radial velocity could be compatible with NGC~1252~18 
      in Table~\ref{t_respec} but both distances and proper motions are incompatible. Finally, TW~Hor and NGC~1252~11 have compatible 
      distance and proper motions but their radial velocities are very different although we cannot confirm that NGC~1252~11 is not a binary. 
      However, TW~Hor and NGC~1252~21 have somewhat compatible (within 6$\sigma$) distance, radial velocity and proper motions. This 
      clearly shows how heterogeneous the stellar populations in this area of sky are and how dangerous is to draw any conclusions without 
      having enough data.

    \subsection{HD 20286}
      HD~20286 is NGC~1252~4 in \citet{Bouchet84}. This star is also known as CD-58 652 and it was identified as a blue straggler candidate 
      in NGC~1252 by \citet{Ahumada95}. \citet{XD05} pointed out that this B9.5 V star is the only confirmed blue straggler in NGC~1252. The 
      object is also included as blue straggler in an updated list by \citet{Ahumada07}. HD~20286 is located 613$^{+407}_{-174}$ pc from the 
      Sun and has ($\mu_{\alpha} \cos\delta, \mu_{\delta}) = (0.93\pm0.63, -4.94\pm0.74)$ mas yr$^{-1}$ \citep{vanLeeuwen07}. Its SPM4
      (1230000090) proper motions are ($\mu_{\alpha} \cos\delta, \mu_{\delta}) = (6.2\pm2.3, -7.1\pm1.9)$ mas yr$^{-1}$. Its UCAC4 
      (161-002935) proper motions are ($\mu_{\alpha} \cos\delta, \mu_{\delta}) = (5.4\pm1.0, -6.3\pm1.0)$ mas yr$^{-1}$. Even if its 
      heliocentric distance could be compatible with our determination for NGC~1252, their proper motions are clearly incompatible. There 
      are no stars in Table~\ref{t_respec} with both distance and proper motions compatible with those of HD~20286. As in the previous case, 
      observational evidence indicates that HD~20286 is not part of NGC~1252. Therefore, its status as blue straggler should be revised. 

    \subsection{HD 20059}
      HD~20059 is NGC~1252~1 in \citet{Bouchet84}. This star is also known as CD-58 641. HD~20059 is located 96$^{+7}_{-6}$ pc from the Sun 
      and has ($\mu_{\alpha} \cos\delta, \mu_{\delta}) = (4.39\pm0.64, -69.81\pm0.78)$ mas yr$^{-1}$ \citep{vanLeeuwen07}. Its SPM4
      (1230000086) proper motions are ($\mu_{\alpha} \cos\delta, \mu_{\delta}) = (-3.2\pm2.7, -63.9\pm2.2)$ mas yr$^{-1}$. The UCAC4 
      (162-002803) proper motions are ($\mu_{\alpha} \cos\delta, \mu_{\delta}) = (6.1\pm1.0, -67.5\pm1.0)$ mas yr$^{-1}$. With a spectral 
      type of K1III/IV, this foreground star is completely unrelated to NGC~1252 and it was already classified as such by \citet{Dias06}. 
      However, it appears to be the primary of a wide binary system with a faint ($V\sim13$) kinematic companion at about 27 arcseconds 
      ($\sim$2,600 AU) from the primary: the star UCAC4 162-002805 has proper motions ($\mu_{\alpha} \cos\delta, \mu_{\delta}) = (8.6\pm2.5, 
      -62.2\pm2.5)$ mas yr$^{-1}$ (SPM4) and ($\mu_{\alpha} \cos\delta, \mu_{\delta}) = (5.8\pm2.8, -64.7\pm2.7)$ mas yr$^{-1}$ (UCAC4).

    \subsection{HD 20037}
      HD~20037 is clearly NGC~1252~13 in \citet{Bouchet84} but SIMBAD does not match their coordinates for unknown reasons. It is also known 
      as CD-58 640. It is located 167$^{+10}_{-9}$ pc from the Sun and has ($\mu_{\alpha} \cos\delta, \mu_{\delta}) = (10.03\pm0.28, 
      4.42\pm0.34)$ mas yr$^{-1}$ \citep{vanLeeuwen07}. Its SPM4 (1230000085) proper motions are ($\mu_{\alpha} \cos\delta, \mu_{\delta}) 
      = (7.4\pm3.8, 2.1\pm3.0)$ mas yr$^{-1}$. The UCAC4 (162-002893) proper motions are ($\mu_{\alpha} \cos\delta, \mu_{\delta}) = 
      (10.0\pm1.0, 4.4\pm1.0)$ mas yr$^{-1}$. With a spectral type of G8III, this foreground star is not related to NGC~1252 although it is
      the closest projected bright star to the cluster (see Fig. \ref{ngc1252field}).

    \subsection{NGC 1252 21}
      NGC~1252~21 has SPM4 (1230015481) proper motions ($\mu_{\alpha} \cos\delta, \mu_{\delta}) = (13.4\pm1.7, 7.1\pm1.6)$ mas yr$^{-1}$.
      Its UCAC4 (162-002824) proper motions are ($\mu_{\alpha} \cos\delta, \mu_{\delta}) = (15.9\pm1.2, 7.4\pm1.2)$ mas yr$^{-1}$. It was 
      considered a probable member of NGC~1252 by \citet{Dias06} with a membership probability of 62 per cent (the highest being 78 per cent). Our results 
      indicate that it is a foreground star of almost solar metallicity located at a distance of 360$\pm$40 pc from the Sun (see Table 
      \ref{t_respec}). Its kinematic properties and distance are compatible with those of TW~Hor. The proper motions of this star match 
      those attributed to NGC~1252 in \citet{Dias06} and \citet{ZPBML12}. 

    \subsection{CD-58 642}
      CD-58 642 is NGC~1252~11 in \citet{Bouchet84}, a K0III star. It was also considered a probable member of NGC~1252 by \citet{Dias06} 
      with a membership probability of 62 per cent. However, our results suggest that it is another foreground star of solar metallicity located at 
      a heliocentric distance of 380$\pm$100 pc. Although its SPM4 (1230000671) proper motions are ($\mu_{\alpha} \cos\delta, \mu_{\delta}) 
      = (17.4\pm1.5, 15.9\pm1.3)$ mas yr$^{-1}$ and its UCAC4 (162-002816) proper motions are ($\mu_{\alpha} \cos\delta, \mu_{\delta}) = 
      (17.4\pm0.9, 16.2\pm0.8)$ mas yr$^{-1}$, compatible with both TW~Hor and NGC~1252~21, its radial velocity is very different, 
      -7.1$\pm$0.6 km s$^{-1}$ but it may be a binary.

    \subsection{NGC 1252 17}
      NGC~1252~17 has SPM4 (1230015463) proper motions ($\mu_{\alpha} \cos\delta, \mu_{\delta}) = (7.3\pm1.4, -17.5\pm1.3)$ mas yr$^{-1}$ 
      and UCAC4 (162-002812) proper motions ($\mu_{\alpha} \cos\delta, \mu_{\delta}) = (7.9\pm1.0, -8.4\pm1.1)$ mas yr$^{-1}$. Our results 
      suggest that it belongs to the background Population II (see Table \ref{t_respec}) and it is located 6000$\pm$500 pc from the Sun with 
      [Fe/H] = -1.15. \citet{Dias06} gave a membership probability of 19 per cent for this object so it is unlikely related to the cluster for them 
      too. With a radial velocity of 67.7$\pm$0.5 km s$^{-1}$ and non-neglibible proper motions, the velocity of this metal poor, halo star 
      may be close to the Galactic escape velocity. Following the procedure outlined at the end of Section 3, we derive the heliocentric 
      reference frame velocity components UVW, which turn out to be U = 305$\pm$55 km s$^{-1}$, V = -380$\pm$40 km s$^{-1}$ and W = 
      239$\pm$33 km s$^{-1}$. This yields a total velocity V$_T$ = 543 km s$^{-1}$. Taking into account the values of the Solar Motion and
      the in-plane circular motion of the local standard of rest around the Galactic Centre from \citet{SBD10}, the Galactocentric velocity 
      components are U = -294$\pm$55 km s$^{-1}$ (U is now positive in the direction of the Sun), V = -125$\pm$40 km s$^{-1}$ and W = 
      246$\pm$33 km s$^{-1}$. NGC~1252~17 is moving in a retrograde orbit around the centre of the Galaxy as it is likely a bound star 
      because its current speed is below the local escape speed \citep[498 $< v_{\rm esc} <$ 608 km s$^{-1}$,][]{Smith07}. On the issue of 
      the origin of this star, we can speculate that it was formed long ago in a globular cluster. Its radial velocity is well in excess of 
      typical values of the globular cluster escape velocity. NGC~1252~17 may have been ejected during interactions with hard binaries at 
      the cluster core, becoming a so-called cannonball star \citep{MDM91}.

\end{document}